\documentclass[twoside,twocolumn,9pt]{article}
\usepackage{extsizes}
\usepackage[super,sort&compress,comma]{natbib} 
\usepackage[version=3]{mhchem}
\usepackage[left=1.5cm, right=1.5cm, top=1.785cm, bottom=2.0cm]{geometry}
\usepackage{balance}
\usepackage{mathptmx}
\usepackage{sectsty}
\usepackage{graphicx} 
\usepackage{lastpage}
\usepackage[format=plain,justification=justified,singlelinecheck=false,font={stretch=1.125,small,sf},labelfont=bf,labelsep=space]{caption}
\usepackage{float}
\usepackage{fancyhdr}
\usepackage{fnpos}
\usepackage[english]{babel}
\addto{\captionsenglish}{%
  
}
\usepackage{array}
\usepackage{droidsans}
\usepackage{charter}
\usepackage[T1]{fontenc}
\usepackage[usenames,dvipsnames]{xcolor}
\usepackage{setspace}
\usepackage[compact]{titlesec}
\usepackage{hyperref}
\usepackage{cases}
\usepackage{amsmath,amsfonts,amssymb} 
\usepackage{mathrsfs,bm}              
\usepackage[normalem]{ulem}           

\usepackage{epstopdf}

\newcommand{\kbT}{k_{\scriptscriptstyle\textrm{B}}T}

\renewcommand{\vec}[1]{\boldsymbol{#1}}
\definecolor{cream}{RGB}{222,217,201}

\begin{document}
\pagestyle{fancy}
\thispagestyle{plain}
\fancypagestyle{plain}{
\renewcommand{\headrulewidth}{0pt}
}

\makeFNbottom
\makeatletter
\renewcommand\LARGE{\@setfontsize\LARGE{15pt}{17}}
\renewcommand\Large{\@setfontsize\Large{12pt}{14}}
\renewcommand\large{\@setfontsize\large{10pt}{12}}
\renewcommand\footnotesize{\@setfontsize\footnotesize{7pt}{10}}
\makeatother

\renewcommand{\thefootnote}{\fnsymbol{footnote}}
\renewcommand\footnoterule{\vspace*{1pt}%
\color{cream}\hrule width 3.5in height 0.4pt \color{black}\vspace*{5pt}} 
\setcounter{secnumdepth}{5}

\makeatletter 
\renewcommand\@biblabel[1]{#1}            
\renewcommand\@makefntext[1]%
{\noindent\makebox[0pt][r]{\@thefnmark\,}#1}
\makeatother 
\renewcommand{\figurename}{\small{Fig.}~}
\sectionfont{\sffamily\Large}
\subsectionfont{\normalsize}
\subsubsectionfont{\bf}
\setstretch{1.125} 
\setlength{\skip\footins}{0.8cm}
\setlength{\footnotesep}{0.25cm}
\setlength{\jot}{10pt}
\titlespacing*{\section}{0pt}{4pt}{4pt}
\titlespacing*{\subsection}{0pt}{15pt}{1pt}

\fancyfoot{}
\fancyfoot[LO,RE]{\vspace{-7.1pt}\includegraphics[height=9pt]{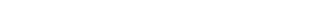}}
\fancyfoot[CO]{\vspace{-7.1pt}\hspace{13.2cm}\includegraphics{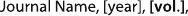}}
\fancyfoot[CE]{\vspace{-7.2pt}\hspace{-14.2cm}\includegraphics{head_foot/RF}}
\fancyfoot[RO]{\footnotesize{\sffamily{1--\pageref{LastPage} ~\textbar  \hspace{2pt}\thepage}}}
\fancyfoot[LE]{\footnotesize{\sffamily{\thepage~\textbar\hspace{3.45cm} 1--\pageref{LastPage}}}}
\fancyhead{}
\renewcommand{\headrulewidth}{0pt} 
\renewcommand{\footrulewidth}{0pt}
\setlength{\arrayrulewidth}{1pt}
\setlength{\columnsep}{6.5mm}
\setlength\bibsep{1pt}

\makeatletter 
\newlength{\figrulesep} 
\setlength{\figrulesep}{0.5\textfloatsep} 

\newcommand{\topfigrule}{\vspace*{-1pt}%
\noindent{\color{cream}\rule[-\figrulesep]{\columnwidth}{1.5pt}} }

\newcommand{\botfigrule}{\vspace*{-2pt}%
\noindent{\color{cream}\rule[\figrulesep]{\columnwidth}{1.5pt}} }

\newcommand{\dblfigrule}{\vspace*{-1pt}%
\noindent{\color{cream}\rule[-\figrulesep]{\textwidth}{1.5pt}} }

\makeatother

\twocolumn[
  \begin{@twocolumnfalse}
{\includegraphics[height=30pt]{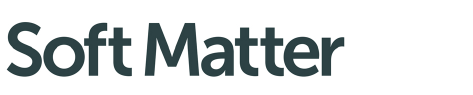}\hfill\raisebox{0pt}[0pt][0pt]{\includegraphics[height=55pt]{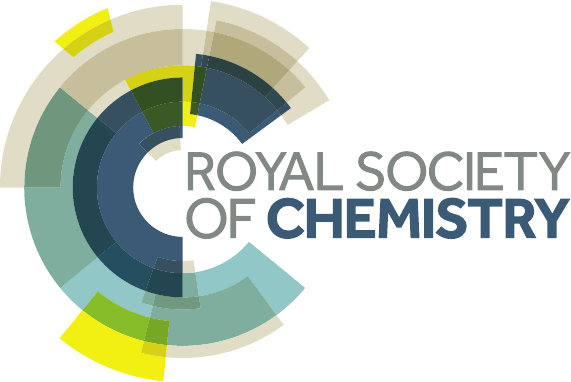}}\\[1ex]
\includegraphics[width=18.5cm]{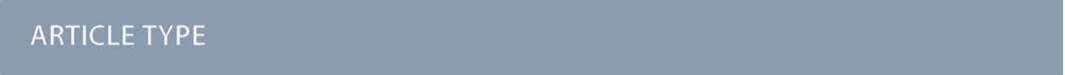}}\par
\vspace{1em}
\sffamily
\begin{tabular}{m{4.5cm} p{13.5cm} }

\includegraphics{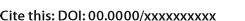} & \noindent\LARGE{\textbf{The roles of elasticity and dimension in liquid-gel phase separation}} \\
\vspace{0.3cm} & \vspace{0.3cm} \\

 & \noindent\large{Shichen Wang,$^{\ast}$\textit{$^{a}$} and Peter D. Olmsted$^{\dag}$\textit{$^{a}$}} \\

\includegraphics{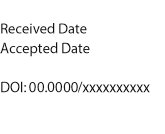} & \noindent\normalsize{We compare six elastic models for polymer networks in the context of phase separation within a gel, including a new model that combines the finite extensible Arruda-Boyce model and the slip tube model for entangled chains. We study incompressible uniaxial stretch and compression, and three volume-changing constrained-dimension deformations, in which the material can only deform in the designated dimensions(s) while the constrained direction(s) remain(s) the same. Each model responds differently to large deformations, and our proposed model successfully describes both  strain softening and strain hardening, which are both present in well-entangled elastomers. When considering phase separation, we show that the commonly-used neo-Hookean model fails to admit a common tangent construction for phase coexistence for 3D deformations. This can be resolved by using a model with finite extension, such as the Arruda-Boyce model. In  constrained-dimension deformations, where the gel's volume is allowed to change, for elastic models in which phase coexistence is possible, the critical temperatures increases and the critical concentration decreases with increasing deformation dimensions. This strong dependence of the phase diagram on spatial dimension and geometry distinguishes phase separation elastic media from conventional phase separation.} \\

\end{tabular}

 \end{@twocolumnfalse} \vspace{0.6cm}
]

\renewcommand*\rmdefault{bch}\normalfont\upshape
\rmfamily
\section*{}
\vspace{-1cm}


\footnotetext{\textit{$^{\ast}$~E-mail: sw1055@georgetown.edu}}
\footnotetext{\textit{$^{\dag}$~E-mail: pdo7@georgetown.edu}}
\footnotetext{\textit{$^{a}$~Department of Physics and Institute of Soft Matter Synthesis and Metrology, Georgetown University, 3700 O St. NW, Washington DC 20057, USA.}}











\section{Introduction}

Phase transitions in polymer gel networks have been studied for decades. The swelling de-swelling phase transition in polymer gels was modeled by Flory and Rehner \cite{10.1063/1.1723792}. The thermodynamics and dynamics of volume phase transitions in gels were extensively explained by Tanaka \cite{Shibayama1993, annurev:/content/journals/10.1146/annurev.ms.22.080192.001331} and Onuki \cite{onuki2002phase,Onuki1993}, as well as the critical behavior and density fluctuation dynamics of such a binary system \cite{PhysRevLett.38.771}. Coarsening and spinodal decomposition in gels were studied by \citet{PhysRevE.59.R1331}. Dimitriyev and co-workers studied gel volume phase transitions with Flory-Rehner model and extreme deformations in gel `tori'  \cite{PhysRevE.98.020501,Dimitriyev_2019}. The Flory-Rehner model was modified to include excluded-volume interactions in order to to model the swelling of heterogeneous microgels \cite{C7CP02434G}. 

An important application of phase separation coupled with elasticity is the formation of membraneless organelles, which can be stabilized by liquid-liquid phase separation\cite{doi:10.1146/annurev-cellbio-100913-013325}. However, this mechanism is less well understood if the cell has a rigid environment, as in the case of condensate (high concentration domains) formation in the nucleus or cytosol \cite{PhysRevLett.126.258102}. Recently, experiments on phase separation in elastic gels, consisting of a polymer network and a solvent, were performed by Dufresne \textit{et al.} \cite{Fernandez-Rico:2024aa,Rosowski:2020ab,D0SM00628A,doi:10.1126/sciadv.aaz0418,PhysRevX.8.011028,Style:2015aa,Jensen14490} to simulate the formation of membraneless organelles in deformable elastic environments. The experiments observed that liquid inclusions were induced in gels by thermal quenching, resulting in a phase separated system where the nucleation temperature \cite{Rosowski:2020ab}, the morphology of the structures \cite{D0SM00628A, Fernandez-Rico:2024aa}, and the characteristic length scale of the structure growth \cite{Fernandez-Rico:2024aa} are all affected by elasticity. 

The main problem regarding phase separation in an elastic environment is the long range and non-additive nature of elasticity. Ordinary phase separation usually assumes two distinct homogeneous regions of composition, separated by a thin interface. Elasticity is a long range interaction in which the strain field caused by the point force decays as $1/R$ \cite{doi:10.1143/JPSJ.58.3069,CAHN1961795}. The long range nature of elasticity suggests that in a phase separated system, instead of two homogeneous region separated by a thin interface, we will have a gradually deformed network and a correspondingly varying concentration field. 
Subsequently, the free energy of the system canbecomes non-additive, losing some thermodynamic qualities. In systems of homogeneous regions separated by sharp interfaces, we can use the common tangent method to find coexisting states \cite{doisoftmatterphys}. 
However, with elasticity, the simple   conditions required in order to use the  common tangent method (extensivity and a localized interface) are usually not satisfied. One solution to circumvent the problem is to use an inhomogeneous model such as the Gent cavity approximation \cite{doi:10.1098/rspa.1959.0016,Horgan:2004aa, HORGAN1992279, Ronceray_2022, KOTHARI2020104153, Zhu:2018aa, D2SM01101H, PhysRevE.107.024418, RAAYAIARDAKANI2019100536}, which describes the inhomogeneous deformation of a cavity surrounded by an elastic matrix, and the energy is calculated by integrating over the entire surrounding matrix.

\begin{figure*}
\centering
\includegraphics[width=0.8\textwidth]{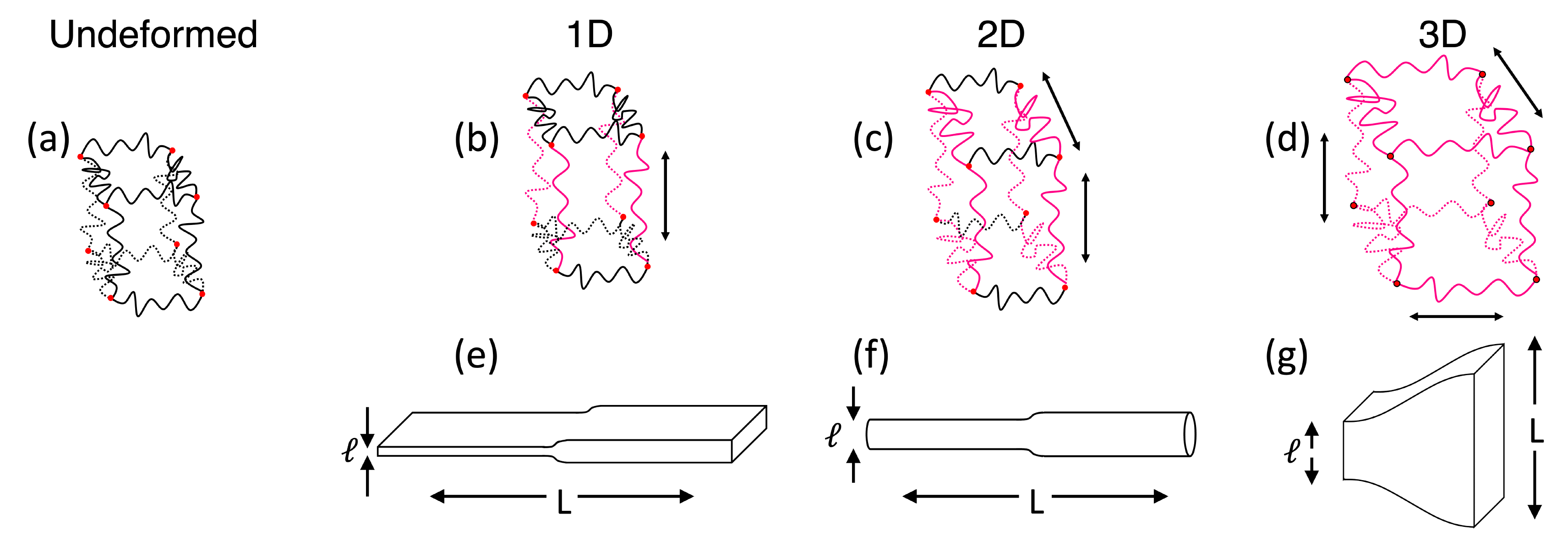}
\caption{Microscopic representations of (a) 1D, (b) 2D, (c) 3D constrained-dimension deformation. Geometry of the material can coincide with the deformation dimension in ways of: (d) 1D deformation of a semi-infinite slab ($L\gg \ell$) (e) 2D deformation of a semi-infinite rod ($L\gg \ell$) and (f) 3D deformation of an isotropic bulk material. }
\label{fig:cddsc}
\end{figure*}

Another factor unique to elasticity is the role of the deformation geometry. In this paper, we will consider two modes of uniaxial deformation and three modes of constrained-dimension deformation. The constrained-dimension deformation describes a compressible elastic material whose deformation is constrained in certain principal axis/axes. Consider an elastic cube and an apparatus that can hold the cube by any one or two of its three pairs of opposing surfaces. No deformation would occur in the direction(s) normal to the surfaces that are held, while in the D direction(s) not restricted by the apparatus the cube is allowed to deform freely by an extension ratio $\lambda$. We then call the deformation of the cube by $\lambda$ in D directions (and constrained to not deform in the other dimensions) a \textit{D-dimensional deformation} (Fig.~\ref{fig:cddsc}bcd). The other dimension of interest is that of the material itself. We can call a material 1$d$ if the length of one side of its geometry is significantly shorter than the other two (Fig.~\ref{fig:cddsc}e), or 2$d$ if two sides are significantly shorter than one (Fig.~\ref{fig:cddsc}f), or 3$d$ when all three sides have comparable length (Fig.~\ref{fig:cddsc}g). 
In this schematic, the dimensions of the material and the deformation coincide, but they are not necessarily bound. For instance, in the example of the cube, which is a 3$d$ material, we can still apply a 1D or 2D deformation. Since the relevant concept in the context of elasticity is the deformation dimension, we will use dimension to indicate the dimensional of the deformation unless it is specified otherwise. 

In this paper, we investigate six elastic models in five modes of deformation to find the best one to characterize experiments similar to those of Refs.~\cite{Fernandez-Rico:2024aa,Rosowski:2020ab,D0SM00628A,doi:10.1126/sciadv.aaz0418,PhysRevX.8.011028,Style:2015aa,Jensen14490}. We show that the elastic free energy density and stress change qualitatively when the dimension of deformation is changed, and that all 6 models have distinct signatures in the different non-linear deformation modes considered. We show that the commonly-used Neo-Hookean model does not admit the familiar common tangent construction in three dimensions. We construct model phase diagrams, and we  illustrate the effects of an often-neglected `volume' contribution to rubber elasticity that arises from the statistical mechanics of rubber \cite{1951JChPh..19.1435W,10.1063/1.1723621}.

\section{Elastic Models}\label{sec:energystress}
\subsection{Overview and parameters}
In this study, we compare five existing elastic models and a new model. The Neo-Hookean model is the `standard' model for cross-linked rubber, which works well for moderate finite deformations \cite{10.1063/1.1723785, Flory:1985aa}. It is well known that the stress-strain response of some rubbers can display stress softening \cite{10.5254/1.3539210}, which has been attributed to the effects of entanglements. This has been addressed by the phenomenological Mooney-Rivlin model \cite{10.1063/1.1712836,doi:10.1098/rsta.1948.0024,BERGSTROM2015209} and the microscopic slip tube model \cite{Rubinstein:2002aa}. Because phase separation into a dilute phase of drops of orders of microns in sizes \cite{PhysRevX.8.011028} will require large stretching, likely to exceed the regime of the NH model and incur strain hardening, we include two finite extensibility models: the Arruda-Boyce model \cite{ARRUDA1993389} and the finite extensible nonlinear elastic (FENE) model \cite{KEUNINGS199785}. Finally, we combine the slip tube model and the Arruda-Boyce model to account for both strain softening and strain hardening.

We define the extension ratio $\lambda_i = \frac{R_i}{R_{0i}}$, where $R$ is the coordinate in the deformed body and $R_0$ in the undeformed body. For a homogeneously deformed body, the deformation gradients $F_{ij} = \frac{\partial R_i}{\partial R_{0j}}$ are indistinguishable from the extension ratios, so that
\begin{equation} \label{eq:dg-1} 
\mathbf{F} =
\begin{pmatrix}
\lambda_x & 0 & 0 \\
0 & \lambda_y & 0\\
0 & 0& \lambda_z 
\end{pmatrix}.
\end{equation}
The deformation gradient $\mathbf{F}$ is then used to define the Finger tensor $W_{ij}= \sum_{k}F_{ik}F_{jk}$. For homogeneous deformations, the invariants of the Finger tensor can be written as 
\begin{subequations}
\begin{align}
I_1 &= \text{Tr} \mathbf{W} = \sum_{i}\lambda_i^2, \label{eq:invi1}\\
I_2 &= \frac{1}{2}\left(I_1^2 - \text{Tr} \mathbf{W}^2 \right) = \sum_{i \neq j} \lambda_i^2 \lambda_j^2,\label{eq:invi2}\\
I_3 &= \det \mathbf{W} = \prod_i \lambda_i^2. \label{eq:invi3}
\end{align}
\end{subequations}
An important feature of the invariant $I_3$ is that it parameterizes the change in volume enclosed by the elastic material, 
\begin{equation} \label{eq:vI3}
\frac{V}{V_0} = \sqrt{I_3}. 
\end{equation} 

We divide the elastic free energy of an elastic network into two parts: an elastic part $f_\text{cf}$ associated with the (conformational) deformation of the crosslinked strands, and a bulk free energy part associated with liquid-like intermonomer interactions that govern volume changes:
\begin{align} \label{eq:fe2p}
\mathcal{F} &= \int dV_0 \left[f^0_{\text{cf}}  + f^0_{\textrm{bulk}}\right]\\
&=\int dV_0 \left[f^0_{\text{cf}}  +
  \tfrac12K(I_3-1)^2\right]. 
\end{align}
where the second term controls the volume change and $K$ is the bulk modulus. Compressibility in a rubber is determined by intermonomer forces, which are much larger than the elastic forces caused by the conformational energy change of the strands, characterized by the strands' end-to-end distances. In this scenario, we can treat the dry rubber as incompressible with an infinite bulk modulus $K\rightarrow\infty$ (see  discussion in Section \ref{subsec:prepro}). If the network is made compressible by allowing solvent particles to occupy or vacate the spaces between polymer molecules, the bulk modulus $K$ is replaced by an elastic contribution to the osmotic modulus, which we will introduce in Section~\ref{sec:dimension}.

The free energy density can be written per unit volume of the reference state ($f^0$) or per unit volume in the current state ($f$) after deformation, as follows: 
\begin{align} \label{eq:fref-fcur}
    {\cal F}&=\int\,dV_0\,f^0 =\int\,dV f =\int\,dV \frac{1}{\sqrt{I_3}}f^0.
\end{align}
\subsection{The Neo-Hookean model}
The Neo-Hookean (NH) model is used predominantly in rubber elasticity. There are multiple types of Neo-Hookean models; Here we choose one in which the network is made of Gaussian chains\cite{TreloarLRG:rubber}, while chain end fluctuations are determined by network functionality $\Phi$, \textit{i.e.} number of chains per crosslinks. The corresponding elastic free energy density is 
\begin{subequations}  \label{eq:n-h-g}
\begin{align}
 f^0_{\textrm{NH}}&= \frac{\nu}{2} C k_B T \left(I_1 -3 \right) = \frac{\Phi}{4} C n_{x0}k_B T \left(I_1 -3 \right),\\
&=  
\tfrac12 n_{x0}k_B T 
\left(I_1 -3 \right) \qquad(\Phi=4),
\end{align}
\end{subequations}
where $\nu$ is the number of polymer strands per unit volume, $n_{x0}$ is the crosslink density in the reference state (undeformed body); and they are related by the relation $n_{x0} = 2\nu/\Phi$; The coefficient $C$ is given by\cite{Flory:1985aa}
\begin{equation} \label{eq:pccoef}
C =1-\frac{2}{\Phi}.
\end{equation}
Since this model characterizes the polymer chains as springs that can be stretched indefinitely, it cannot capture the elastic free energy of large deformations, as may occur during phase separation within a gel. 

\subsection{The Arruda-Boyce model}
The Arruda-Boyce (AB) model \cite{ARRUDA1993389,DAVIDSON20131784} describes the elasticity of freely jointed chains with finite length. The extension cannot exceed the length of the chain, and the AB model more accurately describes elasticity at large deformation. We assume that the effect of junction fluctuations is the same as for Gaussian chains, so that coefficient $C$ remains the same:
\begin{equation} \label{eq:a-b-g}
\begin{split}
f^0_{\text{AB}} = &\frac{\Phi}{2} n_{x0}k_BTC\left[\sqrt{N}\beta\sqrt{\frac{I_1}{3}} - 
 N \ln \left(\frac{\sinh \beta}{\beta}\right)\right], 
 \end{split}
\end{equation}
where $N$ is the number of Kuhn steps in a network strand, and $\beta$ is the inverse Langevin function, which satisfies
\begin{equation} \label{eq:langevin}
\coth \beta - \frac{1}{\beta} = \frac{R}{Nb}=\sqrt{\frac{I_1}{3N}}.
\end{equation}
$N$ is also related to crosslink density $n_{x0}$, mesh volume $\text{v}_{x0}$ and mesh length $\xi$ by $\text{v}_{x0}=R_0^3\equiv\xi^3$:\
\begin{subequations}
\begin{align} \label{eq:nx0toN}
(n_{x0})^{-1} &= \text{v}_{x0} = R_0^3 = (Na^2)^{3/2}\\
\xi &= N^{1/2}a
\end{align}
\end{subequations}
where $a$ is the Kuhn length. 
\subsection{The finite extensible non-linear model}
The (F)inite (E)xtensible (N)on-linear (E)lastic (FENE) model \cite{Peterlin+1966+563+586,Wiest:1989aa, KEUNINGS199785,10.1122/1.3059429} is motivated by the Langevin statistics of a freely jointed chain, yet has a simpler mathematical form than the Arruda-Boyce model:
\begin{equation} \label{eq:fene-g} 
f^0_{\text{FENE}} = \frac{\Phi}{4} n_{x0} k_B T C\left[-3N\text{ln}\left(1-\frac{I_1}{3N}\right)\right].
\end{equation}
\subsection{The Mooney-Rivlin model}
The phenomenological Mooney-Rivlin (MR) model \cite{10.1063/1.1712836,doi:10.1098/rsta.1948.0024,BERGSTROM2015209} was designed to explain the softening effect of an entangled network at moderate deformations $\lambda>1$, where the stress is lower than that predicted by the NH model:
\begin{equation} \label{eq:m-r-g}
f^0_{\text{MR}} = C_1(I_1-3)+C_2(I_2-3) ,
\end{equation}
where $C_1$ and $C_2$ are coefficients extracted to fit the uniaxial stretch data \cite{10.1063/1.434056,YOO20101608}. For $C_2=0$, the MR model reduces to the NH model with $C_1=(\Phi/4)Cn_{x0}\kbT$. This model was designed for uniaxial incompressible extension, where the material is stretched in one direction and contracts in the other two directions. The model can successfully predict the stress for uniaxial stretches, but fails to do so for uniaxial compressions \cite{Rubinstein:2002aa}. 

\subsection{The slip tube model}\label{sec:slip}
Rubinstein and Panyukov proposed the slip tube model \cite{Rubinstein:2002aa} (or equivalently, the RP model) to explain the softening effect of entangled networks in the regime where the MR model fails. The slip tube model is built upon the nonaffine tube model, which considers entanglements as virtual chains attached to the polymer strands. The slip tube model instead allows attachments of the virtual chains to slide along the backbone polymer chains. For isotropic deformations, the slip tube model reduces to the nonaffine tube model since there would not be any slippage. The free energy of the slip tube model is given by 

\begin{equation}
\label{eq:s-t-g} 
\begin{split}f^0_{\text{RP}} &= \frac{\Phi}{4} n_{x0} k_B T \Bigg\{CI_1 + L \sum_{\alpha} \left[\frac{\lambda_{\alpha}}{\sqrt{g_{\alpha}}} + \frac{\sqrt{g_{\alpha}}}{\lambda_{\alpha}}\right] \\
&\qquad\quad-\frac{2L}{3} \ln \left[\left(\frac{N}{L}\right)^3 \prod_{\alpha} g_{\alpha} \right]\Bigg\} ,
\end{split}
\end{equation}
where $g_{\alpha} \equiv N_{\alpha}/N^0_{\alpha}$ are redirection factors defined by the ratio between the number of Kuhn steps in the $\alpha$ direction after the slippage ($N_{\alpha}$) and before the slippage ($N^0_{\alpha}=N/3$), which satisfy the constraint $\sum_{\alpha} g_{\alpha} = 3$ so that $\sum_{\alpha=1}^3 N_{\alpha}=N$. 

The second term in Eq.~\eqref{eq:s-t-g} describes the net effect of a network of entanglements constraining a given chain on a deformation specified by $\lambda_\alpha$ and $g_\alpha$, while the final term is the entropy change to redistributing the monomers of the slipping strands at different lengths $g_{\alpha} N$. For a given set of principal strains $\{\lambda_{\alpha}\}$, the redistribution of monomers $\{g_{\alpha}\}$ is found by minimizing $f^0_{\text{RP}}$ over $g_{\alpha}$ subject to the constraint $\sum_{\alpha}g_{\alpha}=3$, yielding 
\begin{align} 
\begin{split}\label{eq:gtolambda}
&-\frac{\lambda_{\alpha}}{g_{\alpha}^{3/2}} + \frac{1}{\lambda_{\alpha} \sqrt{g_{\alpha}}} -\frac{4}{3} \frac{1}{g_{\alpha}} \\
& = \frac{1}{3} \sum_{\beta} \left[-\frac{\lambda_{\beta}}{g_{\beta}^{3/2}} + \frac{1}{\lambda_{\beta} \sqrt{g_{\beta}}} -\frac{4}{3} \frac{1}{g_{\beta}}\right],
\end{split}
\end{align}
and for a uniaxial extension, $g_\alpha$s are given by \begin{align} 
\begin{split} \label{eq:g-uni}
&-\frac{\lambda}{g^{3/2}}+\frac{1}{\lambda\sqrt{g}} - \frac{4}{3g} \\
&= -\frac{1}{\sqrt{\lambda}}\left(\frac{2}{3-g}\right)^{3/2} + \sqrt{\frac{2\lambda}{3-g}} - \frac{8}{3(3-g)}. 
\end{split}
\end{align}
The number of slip links per chain $L$ is given by:
\begin{equation} \label{eq:st-l}
L = \frac{M_x}{M_e} -1, 
\end{equation}
where $M_e =M_K N$ is the entanglement molecular weight, and $M_K$ is the molecular weight of a Kuhn step. 

For $M_x \leq M_e$, or $N \leq N_e = \frac{M_e}{M_K}$, we expect no entanglements. However, in reality the network mesh size is inhomogeneous, so that even in a system where the average number $N$ of Kuhn steps per strand is less than 
$N_e$
there still exist regions with  $N>N_e$ locally, leading to  entanglements in those regions. In addition, crosslinking is usually a non-equilibrium process that could trap entanglements so that the entanglement spacing is not completely determined by $N_e$. For that reason, $L$ is often treated as a fit parameter \cite{Rubinstein:2002aa, YOO20101608}. In this paper, we assume $N \geq N_e$ for simplicity to demonstrate the RP slip tube model. 

\subsection{The Arruda-Boyce + Rubinstein-Panyukov model}
The elastic models we have discussed so far incorporate either finite extensibility (to capture strain hardening at large $\lambda$) or entanglement (to capture strain softening at intermediate $\lambda$ for entangled networks). By combining finite extensibility with entanglements, a comprehensive model should be able to capture strain softening and subsequent hardening effect \cite{DAVIDSON20131784, Zhu:2018aa}. We propose to combine the Arruda-Boyce model with the slip-tube model (the AB+RP model): 
\begin{align} 
\frac{f^0_{\text{ABRP}}}{ (\Phi/4)n_{x0} k_B T } & = 2 C\sqrt{N}\left[\beta\sqrt{\frac{I_1}{3}} - 
 \sqrt{N} \ln \left(\frac{\sinh \beta}{\beta}\right)\right] \label{eq:abst-g}\\
 & + L \sum_{\alpha} \left[\frac{\lambda_{\alpha}}{\sqrt{g_{\alpha}}} + \frac{\sqrt{g_{\alpha}}}{\lambda_{\alpha}}\right]-\frac{2L}{3} \ln \left[\left(\frac{N}{L}\right)^3 \prod_{\alpha} g_{\alpha} \right]. \nonumber
\end{align}

The main concern of the AB+RP model is that entanglements in the slip tube model were derived by assuming `virtual' Gaussian chains that can slide along the `real' chains. 
While it is tolerable to use Gaussian virtual chains to represent the restraints from entangled freely jointed chains in the small deformation limit, we would underestimate the free energy and stress at larger deformation using the same Gaussian virtual chains when they are stretched beyond the Gaussian limit. Hence, this form of RP model should also incorporate finite extensibility of the virtual chains, which still remains a challenge.
\section{Model Parameterization} \label{sec:para}
From the free energy given by the elastic models, we can find the true elastic stress (force per unit area measured in the deformed reference frame), defined by  $\sigma_{ij} = \delta \mathcal{F}/\delta \varepsilon_{ij}$, where  $\varepsilon_{ij}  = \frac{1}{2} \left[ \frac{\partial u_{i}}{\partial R_{j}} +  \frac{\partial u_{j}}{\partial R_{i}}\right]$ is the infinitesimal strain \cite{LIFSHITZ19861}.
The stress in $\sigma_{ii}$ in the principal direction $i$ is given by
\begin{equation} \label{eq:tstressPA}
\sigma_{ii} = \lambda_{i} \frac{\partial f}{\partial \lambda_{i}},
\end{equation}
where $\lambda_{i}$ is the extension ratio in the $i$ direction. For small deformations, the stress tensor is
\begin{equation} \label{eq:stressmod}
\sigma_{ij} = K\varepsilon_{kk} \delta_{ij} + 2G(\varepsilon_{ij}-\frac{1}{3}\delta_{ij}\varepsilon_{kk}),
\end{equation}
and the elastic energy, to second order in the strain $\varepsilon_{ij} $ is:
\begin{equation} \label{eq:mod}
f_{\text{el}} =  \left[\frac{K}{2} \varepsilon_{ii}^2 + G\left( \text{Tr}\varepsilon^2 - \frac{1}{3} \varepsilon_{ii}^2\right)\right],
\end{equation}
where $G$ and $K$ are, respectively, the shear and bulk modulus.

The most common method used to determine the modulus of a rubber is the incompressible uniaxial tensile test, where a dry piece of elastic material is stretched in one direction by $\lambda$, which measures the Young's modulus $E=\left.\partial(\sigma_{\parallel}-\sigma_{\perp})/\partial(\lambda-1)\right|_{\lambda=1}$.Young's modulus $E=2(1+\nu)G$ is related to shear modulus $G$ and bulk modulus $K$ by Poisson's ratio $\nu = (3K-2G)(6K + 2G)$. For an incompressible material, $K \to \infty$, and Poisson's ratio $ \nu = 1/2$, so $E = 3G$.

\begin{figure}[hbt!]
\centering
\includegraphics[width=0.48\textwidth]{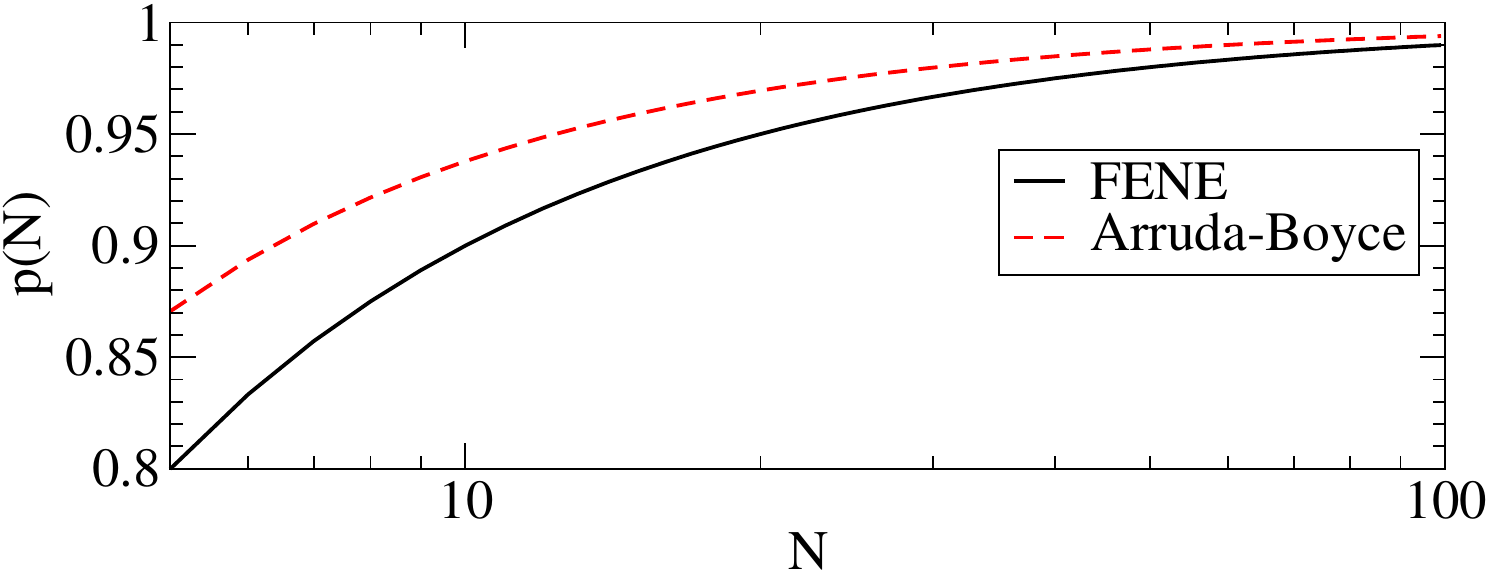}
\caption{Normalization factors $p(N)$ for the finite extensible models. 
}
\label{fig:pnpq}
\end{figure}

By matching the coefficients in Eq.~\eqref{eq:mod} with those of the models, we find:
\begin{subnumcases}{ \frac{E}{3} = \label{eq:nx0nums}}
    \frac{\Phi}{2}C n_{x0} k_B T & (NH) \label{eq:modNH}\\[6truept]
    \frac{\Phi C n_{x0} k_B T}{2 p(N)} & (FENE, AB) \label{eq:modFA}\\[6truept]
    2(C_1+C_2) & (MR) \label{eq:modMR}\\[6truept]
    (2C+\frac{4}{7} L) \frac{\Phi}{4} n_{x0} k_B T & (RP) \label{eq:modRP}\\[6truept]
    \left(\frac{2C}{p(N)}+\frac{4}{7} L\right) \frac{\Phi}{4} n_{x0} k_B T & (AB+RP) \label{eq:modABRP}
\end{subnumcases} 
where $p(N)$ is the normalization factor for the AB and FENE models:  
\begin{subnumcases} {p(N) =}
  \frac{\left.\frac{\partial f^0_{\text{NH}}}{\partial I_1}\right\vert_{I_1 = 3}}{\left.\frac{\partial f^0_{\text{AB}}}{\partial I_1}\right\vert_{I_1 = 3}} = \frac{3}{\sqrt{N} \mathscr{L}^{-1}(1/\sqrt{N})} & (AB)\label{eq:a-b-pn} \\
 \frac{\left.\frac{\partial f^0_{\text{NH}}}{\partial I_1}\right\vert_{I_1 = 3}}{\left.\frac{\partial f^0_{\text{NH}}}{\partial I_1}\right\vert_{I_1 = 3}} = 1-\frac{1}{N}& (FENE)\label{eq:fn-pn} 
 \end{subnumcases}
and $\mathscr{L}^{-1}(1/\sqrt{N})$ is the inverse Langevin function $\beta$ evaluated at $I_1 = 3$, \textit{i.e.}  $\coth \beta-\frac{1}{\beta} = \sqrt{\frac{1}{N}}$. For $N>10$, 
$0.9<p(N)<1$ (Fig.~\ref{fig:pnpq}) and the AB and the FENE model have almost the same modulus $G$ to crosslink density $n_{x0}$ relation as the NH model. We will approximate $p(N)\simeq1$ to simplify the calculations below. 

To compare elastic models, we choose parameters that yield the same shear modulus $G$. As a result,  for the approximation $p(N)\simeq 1$ the crosslink density $n_{x0}$ of the two finite extensibility models (FENE, AB) is the same as that of the Neo-Hookean model, but differs from the MR, RP and AB+RP models. The coefficients $C_1$ and $C_2$ are often reported as the ratio $C_2/C_1$ for a fixed $C_1$ \cite{polc.5070530125, BOYER1987399, 10.1063/1.434056}. The slip tube model has an additional term that involves the number of slip links/entanglements $L$.

We will use PDMS properties to illustrate the differences in the models. PDMS has a Kuhn length $a = 11.3 \textrm{\AA}$ \cite{PhysRevX.12.021047}, a Kuhn molecular weight $M_{\text{Kuhn}} = 381 \textrm{g/mol}$, and an entanglement molecular weight $M_e = 12 \textrm{kg/mol}$ \cite{Fetters:1999aa}. We choose a functionality $\Phi = 4$, which leads to $C= 1/2$. For the NH, AB and FENE models, we can relate the crosslink density $n_{x0}$ to the Young's modulus and $N$ by 
\begin{subequations}
\begin{align} \label{eq:3Mparameter}
n^{\text{NH}}_{x0} = \frac{n^{\text{AB}}_{x0}}{p(N)} =  \frac{n^{\text{FN}}_{x0}}{p(N)} &= \frac{E}{6C k_B T} = \frac{E}{3 k_B T},
\end{align}
\end{subequations}
where we approximate $p(N)\simeq1$.
For the Mooney-Rivlin model,
\begin{equation} \label{eq:MRparameter}
n^{\text{MR}}_{x0}  = \frac{n^{\text{NH}}_{x0}}{1+C_2/C_1}
\end{equation}
where we choose $C_1 = Cn_{x0}k_BT$. An example value for the ratio is 
$C_2/C_1 = 0.58$ \cite{BOYER1987399} for a vulcanized PDMS network with shear modulus $G = 0.2$ MPa. Experiments on end-linked PDMS with $M_x$ = 30 kg/mol (close to the value 28 kg/mol polymer used by Eric Dufresne \cite{PhysRevX.8.011028}) and Young's moduli of 0.07 to 0.4 MPa estimated a ratio of $C_2/C_1$ from 0.78 to 1.6 \cite{YOO20101608}. We observe that the value of $C_2/C_1$ does not qualitatively change the behavior of the model, and that a higher larger value of $C_2/C_1$ softens the stress more in uniaxial extension. We use $C_2/C_1 = 0.58$ for the rest of the paper.

In the NH, AB, and FENE models, $N$ and the mesh length $\xi$ are related to the Young's modulus $E$ by
\begin{subequations}
\label{eq:nxi}
\begin{align}
N^{\text{NH}} &= \left(\frac{6Ck_BT}{Ea^3}\right)^{2/3} = \left(\frac{3k_BT}{Ea^3}\right)^{2/3}, \label{eq:nreNH}\\
\xi^{\text{NH}} & = \text{v}_{x0}^{1/3} = \left(\frac{E}{3 k_B T}\right)^{-1/3}.\label{eq:xiNH}\\
N^{\text{AB}} &= N^{\text{FN}} = \left(\frac{6Ck_BT}{Ep(N)a^3}\right)^{2/3} = \left(\frac{3k_BT}{E\,p(N)a^3}\right)^{2/3}, \label{eq:nreABFN}\\
\xi^{\text{AB}} & = \xi^{\text{FN}} =  \text{v}_{x0}^{1/3} = \left(\frac{E\,p(N)}{3 k_B T}\right)^{-1/3}.\label{eq:xiABFN}
\end{align}
\end{subequations}

To find the relation between $n_{x0}$ and $E$ in the Rubinstein-Panyukov slip tube model, we combine Eqs.~(\ref{eq:nx0toN}, \ref{eq:st-l}, \ref{eq:modRP}) to obtain  
\begin{equation} \label{eq:PRparameter-b}
\frac{14C-4}{7} n^{\text{RP}}_{x0}+ \frac{4}{7} \frac{M_K}{M_e} \frac{(n_{x0}^{\text{RP}})^{1/3}}{a^2}  =  \frac{E}{3 k_B T},
\end{equation} 
where we have assumed that the number of entanglement per chain $L$ is given by Eq.~\eqref{eq:st-l}.
The crosslink density $n_{x0} = 1/(N^{3/2}a^3)$, so that the relation between the modulus and the polymer strand length for the RP model in the `ideal' case (for $C=1/2$) is 
\begin{equation} \label{eq:PRparameter-c}
\frac{3}{7N^{\textrm{RP}}} + \frac{4N^{\textrm{RP}}}{7N_e}  =  \frac{E  a^3 \sqrt{N^{\textrm{RP}}}}{3 k_B T}.
\end{equation} 

For PDMS, $N_e\simeq 12,000/381\simeq32$, so that entanglements should only exist for $N \geq N_e$, which corresponds to a crosslink density $n_{x0} \leq 3.92 \times 10^{24} \textrm{m}^{-3}$, or a Young's modulus 
$E \leq 48.4 \textrm{kPa}$ at $T= 298$ K. 
This value is lower than many reported moduli of PDMS that have shown softening effects (for example, \citet{10.1063/1.434056} found softening for PDMS with moduli E = 27 kPa to 138 kPa). Although the 27 kPa samples satisfy the criterion above, softening at moduli as high as 138 kPa shows the presence of entanglements even when $N<N_e$. As discussed in Section.~\ref{sec:slip}, one potential explanation for this `contradiction' is that the crosslink density is highly inhomogeneous, leading to regions with longer, entangled strands. This hypothesis is supported by the reported polydispersity index of an entangled PDMS (non-zero $C_2$ value) ranging from 1.2 to 2.4 for samples whose chain molecular weight average from $2$ kg$\cdot$mol$^{-1}$ to $58$ kg$\cdot$mol$^{-1}$. In addition, both \citet{10.1063/1.434056} and \citet{PhysRevX.8.011028} used vinyl-terminated polymer strands to synthesize their network, which means that the strands are end-linked. End-linked polymer networks typically have lower polydispersity than randomly crosslinked polymers \cite{PhysRevLett.71.645}. The vulcanized PDMS network would have a higher polydispersity, since crosslinking can occur on any point of the strand, which means that a vulcanized PDMS can easily exceed the maximum modulus we calculated above and still have entanglement present in the network. The parameters discussed in this section are listed in Table~\ref{tab:fig-parameters} for a $E = 40$ kPa end-linked PDMS network.  
\begin{table*}[h]
\centering
\begin{tabular}{c|cccccc}
\hline\hline
$E = 40$ kPa \indent & NH \indent & AB \indent & FN \indent & MR \indent & RP \indent & AB+RP \indent \\ \hline
$10^{3} n_{x0} (\text{nm}^{-3})$ & 3.2 & 3.2 & 3.2 & 2.1 & 2.9 & 2.9  \\[6truept] 
$N$ & 35.8 & 35.8 & 35.8 & 48.5 & 38.9 & 38.9 \\[6truept]
$C_2/C_1$ & - & - & - & 0.58 & - & -\\[6truept]
$N_e$ & - & - & - & - & 37.7 & 37.7\\[6truept]\hline
$\lambda_\text{min}$(Uni) & 0.02 & 0.02 & 0.02 & 0.01 & 0.02 & 0.02\\[6truept]
$\lambda_\text{max}$(Uni) & 10.4 & 10.4 & 10.4 & 12.1 & 10.8 & 10.8\\[6truept]
$\lambda_\text{max}$(1D) & 10.3 & 10.3 & 10.3 & 12.0 & 10.7 & 10.7\\[6truept]
$\lambda_\text{max}$(2D) & 7.29 & 7.29 & 7.29 & 8.50 & 7.60 & 7.60 \\[6truept]
$\lambda_\text{max}$(3D) & 5.98 & 5.98 & 5.98 & 6.96 & 6.24 & 6.24\\[6truept]\hline
$\alpha^*$ (w/ solvent) & 0.50 & 0.517 & 0.529& 1.539 & 0.513 & 0.529 \\[6truept]
$\chi_0(320$K) & 0.8692 & 0.8689 & 0.8687& 0.8589 & 0.8690 & 0.8688\\[6truept]
$\phi_\text{sw}(\chi_0 = 0.87)$ & 0.595 & 0.595 & 0.595& 0.603 & 0.595 & 0.595 \\\hline\hline
\end{tabular}
\caption{Parameters used in figures corresponding to a Young's modulus of $E = 40$ kPa, and the maximum extension $\lambda_{\textrm{max}}$ of different modes of deformation. $n_{x0}$ and $N$ are calculated with Eq.~\eqref{eq:nx0toN} and \eqref{eq:nx0nums}; the number of Kuhn steps between entanglement $N_e$ is for PDMS melts (see  Sec.~\ref{sec:para}). $\lambda_\text{min}$ and $\lambda_\text{max}$ are the extension ratios in the loading direction for uniaxial compression and extension, respectively; $\lambda_\text{max}$ for the constrained-dimension deformation are in the unconstrained direction. $\alpha^{\ast}$ in the third line from bottom was calculated for each model from the swelling equilibrium condition Eq.~\eqref{eq:chi0} for a swollen PDMS whose swelling equilibrium concentration is $\phi_{\text{sw}} = 0.595$ at $\chi_0(320K) = 0.8692$. Alternatively, in the second to last line we assume that $\alpha = 0.5$ and $\phi_\text{sw} = 0.595$ are known, and calculate the value of $\chi_0$ for each model needed to satisfy Eq.~\eqref{eq:chi0}. Finally the last line shows different swelling equilibrium concentrations calculated if we know  $\alpha=0.5$ and $\chi_0 = 0.8692$. We chose to vary $\alpha$ for our calculations in this paper (the row with $\alpha^*$), and the other rows are for comparison. See detailed discussion in Section~\ref{subsec:prepro}.
}
\label{tab:fig-parameters}
\end{table*}

\begin{figure*}[htb!]
\centering
\includegraphics[width=0.9\textwidth, ,trim={0.05cm 0 0 0.05cm},clip]{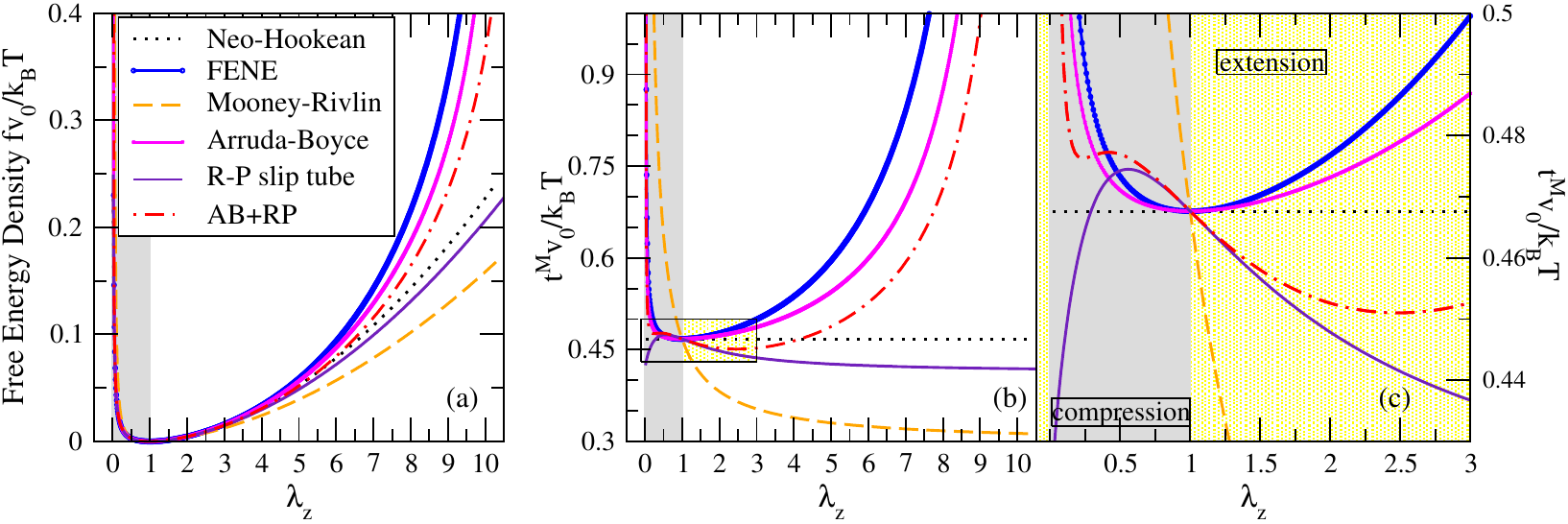}
\caption{Elastic free energy (a) and Mooney ratio (b, c) of an incompressible PDMS network with Young's modulus $E = 40$ kPa undergoing uniaxial tension ($\lambda_z>1$) and compression (($\lambda_z<1$, gray). Parameters are listed in Table.~\ref{tab:fig-parameters}. The Mooney ratio of the NH model is constant as designed (horizontal black dotted line), which represents chains that behave like harmonic springs.}
\label{fig:modelsESb}
\end{figure*}

\section{Models in different modes of deformation}
\subsection{Uniaxial Deformation of Incompressible Materials}
If we apply a uniaxial stretch along the z-axis, the deformation gradient can be written in matrix form as:
\begin{equation} \label{eq:dg-uni} 
\mathbf{F} =
\begin{pmatrix}
\frac{1}{\sqrt{\lambda}} & 0 & 0 \\
0 & \frac{1}{\sqrt{\lambda}} & 0\\
0 & 0& \lambda 
\end{pmatrix}, \\
\end{equation}
and the invariants in terms of the extension ratio $\lambda$ in the $z$ direction become 
\begin{subequations}
\begin{align}
I_1 &= \lambda^2 + \frac{2}{\lambda}, \label{eq:invi1-b}\\
I_2 &=  \frac{1}{\lambda^2} + 2\lambda, \label{eq:invi2-b}\\
I_3 &= 1. \label{eq:invi3-b}
\end{align}
\end{subequations}

We study the uniaxial stretch of both tension ($\lambda_z  \geq 1$) and compression ($\lambda_z \leq 1$). For FENE and AB, when the strand length is sufficiently large, the maximum stretch for tension is $\lambda^{\text{max}}_z \simeq \sqrt{3N}$, and for compression, the maximum stretch in the transverse plane is $\lambda^{\text{max}}_x \simeq \sqrt{\frac{3N}{2}}$, or that the minimum stretch ratio $\lambda_\text{min}$ in the loading direction is $\lambda_z^\text{min} \simeq \sqrt{\frac{2}{3N}}$ (see Appendix~\ref{appx:maxl} for precise limits). The values of maximum extension (for stretch) and minimum extension (for compression) for a PDMS network with $E = 40$ kPa are listed in Table.~\ref{tab:fig-parameters}. Lastly, we present the total tension $t_{zz} = \sigma^{\text{el}}_{zz} - \sigma^{\text{el}}_{xx}$ for the uniaxial deformations \cite{LIFSHITZ19861}. We will adopt the Mooney ratio $t^{M}$ \cite{10.1063/1.1698152,doi:10.1098/rsta.1948.0024} to quantify the stress for a uniaxial deformation: 
\begin{equation}
    t^{M} = \frac{t_{zz}}{\lambda^2-1/\lambda} = \frac{\sigma^{\text{el}}_{zz} - \sigma^{\text{el}}_{xx}}{{\lambda^2-1/\lambda}}.
\end{equation}

\begin{figure}[hbt!]
\centering
\includegraphics[width=0.4\textwidth,trim={0.1cm 0 0 0.05cm},clip]{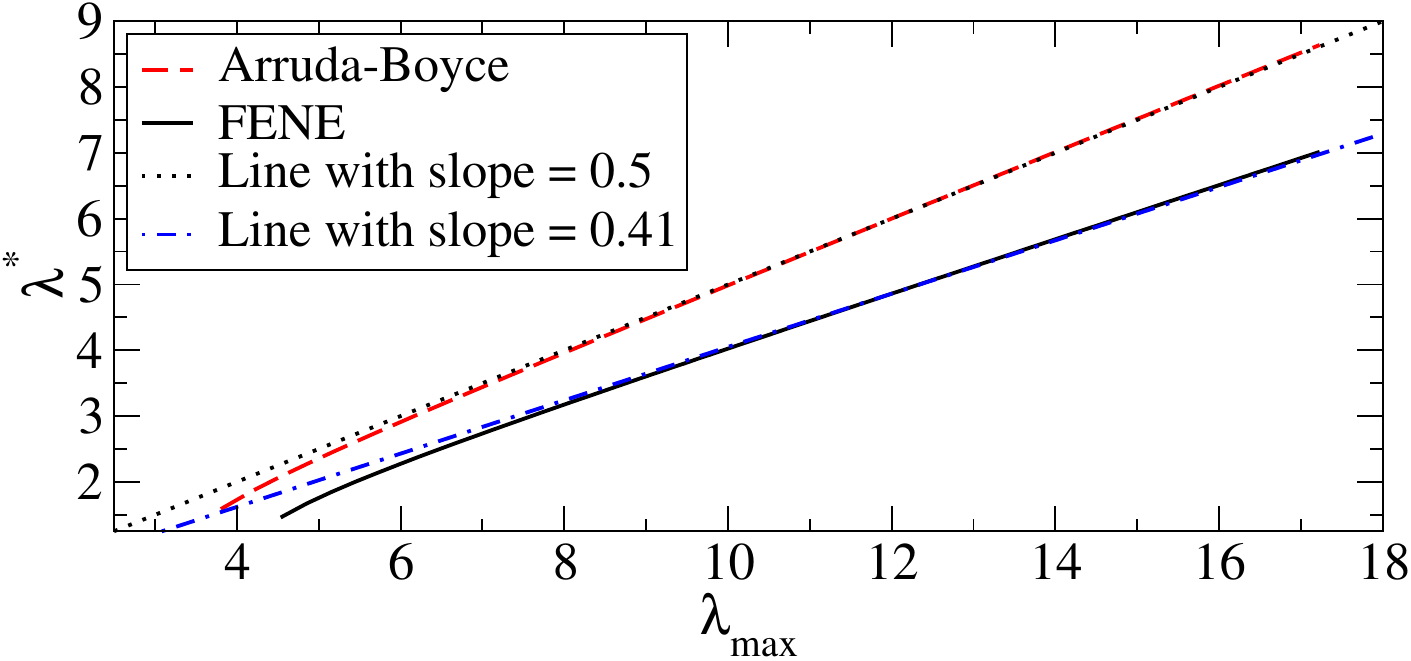}
\caption{Extension ratio upper limit $\lambda^*$ of the linear regimes for the finite extensible models FENE and AB. When the Mooney ratio of the AB and FENE models is $120\%$ that of the NH model, $\lambda^{\ast} \simeq 0.5\:\lambda_\text{max}$ for the AB model and $\lambda^{\ast} \simeq 0.405\:\lambda_\text{max}$ for the FENE model.}
\label{fig:lstar}
\end{figure}

The range of the validity of the Neo-Hookean model can be estimated by finding the stretch ratio $\lambda^{\ast}$ with which the stress given by the AB or the FENE model exceeds that of the Neo-Hookean model by, for example, no more than 20\%, \textit{i.e.} $t^M_\text{AB/FENE}(\lambda^*) = 1.2\, t^M_\text{NH}(\lambda^*).$
As shown in Fig.~\ref{fig:lstar}, $\lambda^*$ grows roughly linearly with $\lambda_\text{max}$ for $\lambda\geq6$ ($N\geq12$), with $\lambda^* \simeq 0.5\:\lambda_\text{max}$ for the AB model and $\lambda^* \simeq 0.41\:\lambda_\text{max}$ for the FENE model. 

Fig.~\ref{fig:modelsESb} shows the free energy density and the Mooney ratio $t^{M}$ for the uniaxial deformations of all models, including both uniaxial extension $(\lambda_z>1)$ and uniaxial compression ($\lambda_z<1$, often called equibiaxial extension). We parametrized the models with the same elastic modulus so that they converge near $\lambda = 1$. The linear stress-strain response should continue at larger stretch for the NH model, while the entanglement models (MR and the slip tube) should show strain softening effect and the finite extensible models (AB and FENE) strain hardening effects. The combined AB-RP model should show both the softening effect and the hardening effect. 

As shown in Fig.~\ref{fig:modelsESb}a, the free energy of the NH model (dotted black line) separates the strain softening and strain hardening models in stretching. It is easier to see in Mooney plots in Fig.~\ref{fig:modelsESb}bc that, while the Mooney ratio of the NH model remains constant, the slip tube model shows strain softening in stretch, and a higher stress in compression near $\lambda = 1$ but eventually decreases below the NH model for large compression $\lambda\ll1$, which agrees with previous studies \cite{Rubinstein:2002aa,DAVIDSON20131784}. There has been little discussion of the nonmonotonic behavior of the slip tube model under compression in the literature, even though the peak of the Mooney ratio has long been observed \cite{YOO20101608, doi:10.1021/acs.macromol.8b02485}. Mathematically, the peak is caused because the Mooney ratio for the slip tube model always has zero slope at $\lambda = 0.56$, independent of the entanglement density $n_{x0}L$ (see Supplementary Information Sec.~1.1 for details). Previous studies showed that if there is no slippage of the entanglements (the non-affine tube model), \textit{i.e.} $g = 1$, the stress remains higher than the NH model in compression \cite{Rubinstein:2002aa}, indicating that slippage of entangled chains causes the stress to decrease. In other words, the deformation of the entangled chains (parameterized by $\lambda$) contributes to increasing the stress (\textit{i.e.} `compression hardening'), but the subsequent slippage of the entanglements (parameterized by $g$) causes the stress to decrease. Physically, it seems that it costs more free energy to deform entangled chains than crosslinked chains under moderate uniaxial compression. This is also evident upon comparing the tube model with the FENE and AB models for $0.5<\lambda<1$. The FENE and the AB models by design have stiffer freely jointed chains but predict lower stresses than the slip tube model, indicating that `compression hardening' is likely caused by the entangled strands. \citet{Rubinstein:2002aa} compared his model with the experiment and found that the slip tube model fits the stress curve the best for both uniaxial compression and tension.

On the other hand, the Mooney-Rivlin model, devised to describe the uniaxial stretch, poorly describes the elasticity of uniaxially compressed polymer networks \cite{Rubinstein:2002aa}. The two finite extensible models (FENE and AB) exhibit similar behavior for both uniaxial tension and compression, whereas AB is slightly less stiff, having lower free energy and stress at large deformations. For a small deformation (Fig.~\ref{fig:modelsESb}c), the AB+RP model has a Mooney ratio curve that closely resembles the original slip tube model, but with slightly higher stress. This is expected because in this range ($\lambda < 0.5\lambda_\text{max} $), the strain hardening effect of the AB model is still negligible, so the softening effect of the slip tube part of the model is more prominent. At around half of the maximum stretch, the AB part of the model takes over and the strain hardening effect overwhelms the softening effect. Although the AB+RP model does not accurately depict the microscopic structure network, it still exhibits the essential features of the models it comprises, namely softening past the initial linear regime and hardening when approaching the maximum stretch. 

\subsection{Constrained-dimension swelling}\label{sec:dimension}

Swelling in a compressible polymer network involves volume changes (of the enclosed space, not the occupied space) by exchanging solvent with the solvent reservoir, which usually involves an isotropic deformation. Here, we incorporate anisotropic deformations during swelling such that the deformation is $\textrm{D}=1$, 2, or 3-dimensional. When the volume changes, solvent particles enter or exit the space, so that the (essentially) infinite bulk modulus is replaced by an osmotic contribution from the crosslink fluctuations, such that the elastic energy in Eq.~(\ref{eq:fe2p}) becomes:
\begin{equation} \label{eq:fingel}
f^0_{\text{el}} = f^0_{\text{cf}} - \frac{\Phi}{4}n_{x0}k_B T \alpha \ln{I_3},
\end{equation}
with the entropic change of the crosslink fluctuation $(\Phi/4) n_{x0}k_B T \alpha \ln{I_3}$ substituting the bulk modulus $K/2$ term. Flory initially included this term with $\alpha = \frac{1}{C\Phi} = \frac{1}{2}$ (for $\Phi = 4$ and $C = 1/2$)  \cite{1951JChPh..19.1435W,10.1063/1.1723621} under the assumption that the space that the crosslinks can explore deforms affinely with the network. \citet{doi:10.1098/rsta.1976.0001} argued that $\ln I_3$ should be generalized to $n_{x0}k_B T \ln \prod_i \omega_i^{-1/2}$, where $\omega_i$ is the deformation associated with the volume explored by the crosslinks. They argued that in a crosslink-dominated gel, $\omega_i$ is constant, independent of $\lambda_i$, while in a highly entangled gel $\boldsymbol{\omega} (\boldsymbol{\lambda})$ depends on the network strain $\boldsymbol{\lambda}$. \citet{10.1063/1.1699106} argues that $\alpha=0$; \citet{PANYUKOV19961} agrees that $\alpha=0$ using replica field theory. \citet{Carla_unpublished_work} argues that the value $\alpha=1/2$ gives rise to a stress-free condition at $\lambda = 1$, which, coincidentally, agrees with Flory's value.

\citet{Onuki1993} and \citet{Muthukumar_2023} considered D $=1$ and D $=2$ swelling with $\lambda_\perp$ fixed and variable $\lambda_\parallel$ (D $=1$); or fixed $\lambda_{\parallel}$ and variable $\lambda_{\perp}$ (D $=2$). Here, we similarly simplify the geometries by considering a compressible polymer network that swells by allowing deformation in certain directions while keeping other dimensions undeformed; \textit{i.e.} constrained $\lambda_{\perp1} = \lambda_{\perp2} = 1$ and variable $\lambda_\parallel = \lambda$ for 1D, or $\lambda_{\parallel} = 1$ and $\lambda_{\perp1} = \lambda_{\perp2} = \lambda$ for D $=2$. For example, a sheet of gel (a quasi 2-dimensional material) can undergo a 1D deformation to thicken (in the transverse direction, Fig.~\ref{fig:cddsc}e). Similarly, a rod (a quasi 1-dimensional material) can stretch radially in 2D (Fig.~\ref{fig:cddsc}f). Finally, a block of gel can swell isotropically in 3D (Fig.~\ref{fig:cddsc}g). The principal stretch ratios in these cases are given by
\begin{align}\label{eq:lambdad}
\vec{\lambda}=\begin{cases}
(\lambda,1,1)& (\text{D}=1) \\
(\lambda,\lambda,1)& (\text{D}=2) \\
(\lambda,\lambda,\lambda)& (\text{D}=3).
\end{cases}
\end{align}
The Finger tensor $W_{ij}$ for these deformations can be written as
\begin{align}
\mathbf{W}^{(\text{D})} &=
\begin{pmatrix}
\lambda^{(\text{D}-1)(\text{D}-2)} & 0 & 0 \\
0 & \lambda^{(\text{D}-1)(4-\text{D})} & 0\\
0 & 0& \lambda^2 
\end{pmatrix} \label{eq:wD}
\end{align}

The invariants of the deformation tensor for a D$-$dimension deformation become 
\begin{subequations}
\begin{align}
I_1 &= \text{D}\lambda^2 + 3-\text{D}, \label{eq:invi1-c}\\
I_2 &= \frac{\text{D}^2-\text{D}}{2} \lambda^4 +\text{D}(3-\text{D})\lambda^2+\frac{(\text{D}-3)(\text{D}-2)}{2}, \label{eq:invi2-c}\\
I_3 &= \lambda^{2\text{D}}. \label{eq:invi3-c}
\end{align}
\end{subequations}

\begin{figure*}[htb!]
\centering
\includegraphics[width=0.9\textwidth]{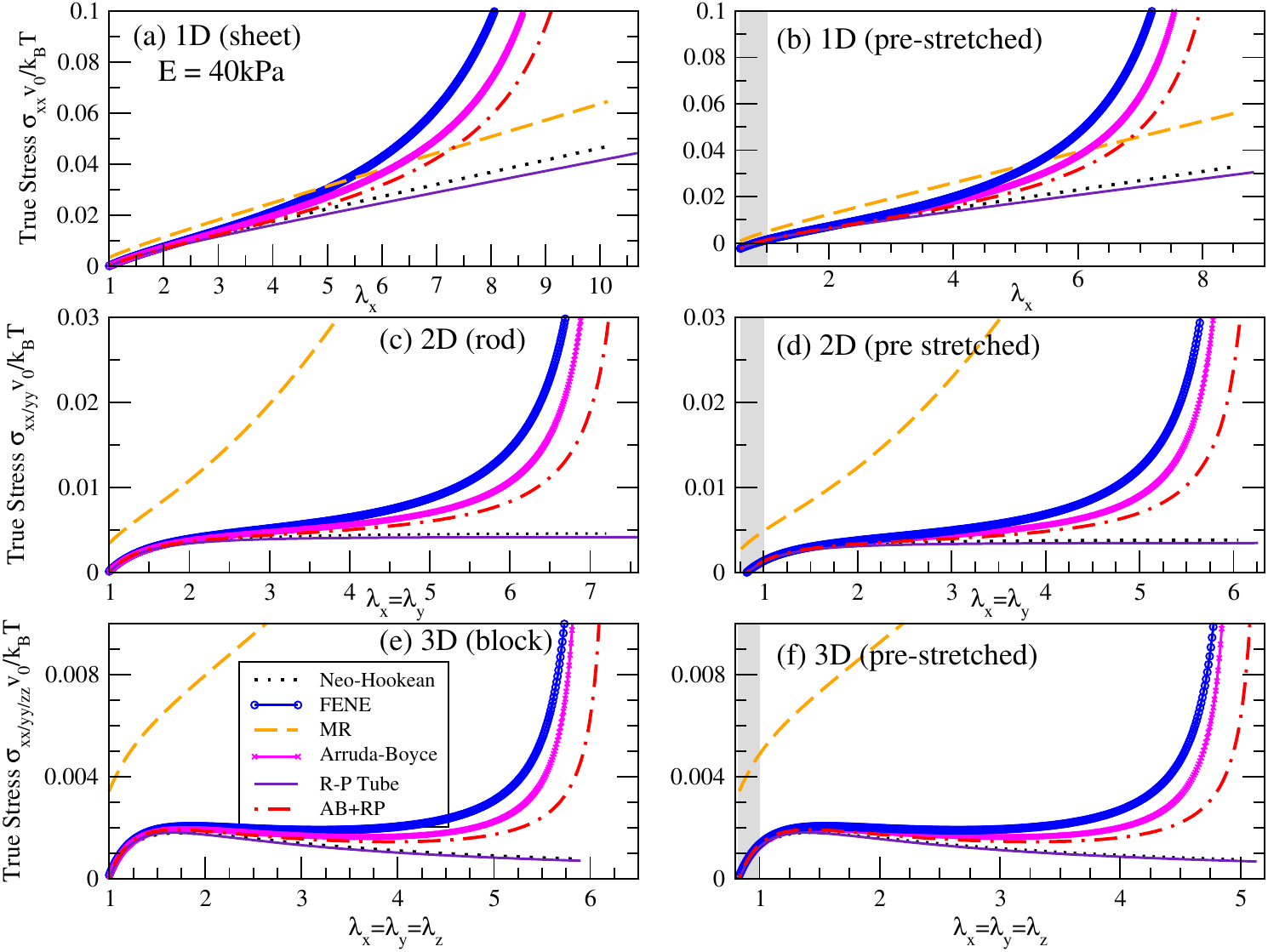}
\caption{True elastic stresses of a dry network swelling in (a) 1D (c) 2D and (e) 3D; and the elastic stresses of a pre-stretched network with $\lambda_0 = 1.2$ shrinking ($\lambda<1$, in gray) and swelling ($\lambda>1$) in (b) 1D (d) 2D and (f) 3D.} \label{fig:EScompressible}
\end{figure*}

Fig.~\ref{fig:EScompressible} shows the true elastic stresses of the six models in 1, 2, and 3D deformations, swelling from a dry state (Fig.~\ref{fig:EScompressible}ace) or from a pre-stretched state (Fig.~\ref{fig:EScompressible}bdf).
We use the parameters listed in Table~\ref{tab:fig-parameters}, assuming the same incompressible Young's modulus for all materials (corresponding to a dry polymer network). The AB, FENE, and ABRP models have a maximum stretch that depends on the dimension of the deformation according to $I_1(\lambda_{\textrm{max}})=3N$, leading to $\lambda_\text{max} = \sqrt{(3N-3+\text{D})/\text{D}}$.

Fig.~\ref{fig:EScompressible}ace shows the elastic stresses of the network swelling from a dry state ($\lambda = 1$ in all directions). We observe that the MR model does not show softening for any constrained-dimensional swelling, because the second invariant $I_2$ depends on $\lambda^2$ and $\lambda^4$, in contrast to the incompressible case (compare Eqs.~\ref{eq:invi2-b} and \ref{eq:invi2-c}). In contrast, the slip tube model exhibits softening for all constrained-dimensional swelling. The two finite extensible models FENE and AB have higher stresses than the NH model, and similarly to the uniaxial stretch, the FENE model has a slightly higher stress than the AB model. The AB+RP model does not exhibit an obvious softening effect compared to the NH model, since its stress is very close to the NH model in swelling when $\lambda$ is close to 1 and becomes higher than the NH model at relatively small $\lambda$ (compared to the uniaxial stretch where the same occurs around $\lambda \approx 0.5\lambda_\text{max}$). 
It is also worth noting that the true stress of the NH model decreases with increasing stretch ratio $\lambda$ for 3D deformation. This is because although the force exerted by and on the polymer strand would continue to increase with extension, the unit surface area against which we measure the stresses would increase faster with a higher-dimension deformation. The stresses of the finite extensible models also decrease, but would eventually be overcome by the divergence in the FENE elasticity.

We emphasize that with the deformation characterized by the Finger tensor of Eq.~(\ref{eq:wD}), the network can only expand ($\lambda \geq 1$) because the dry network is incompressible. For that reason, in order to observe the gel shrinking, we need to add an isotropic `pre-stretching' $\lambda_0$ 
so that the total deformation gradient is given by $\mathbf{F} = \mathbf{F}^{(\text{D})}\cdot \mathbf{F}^{0}$, where $\mathbf{F}^{(\text{D})}$ is the deformation gradient of the \text{D}-dimension with the stretch ratio $\lambda = R/R_\text{ps}$ between the current position $R$ and the pre-stretched position $R_\text{ps}$; and 
\begin{align}
\mathbf{F}^{0} &=
\begin{pmatrix}
\lambda_0 & 0 & 0 \\
0 & \lambda_0 & 0\\
0 & 0& \lambda_0 
\end{pmatrix}. \label{eq:F0}
\end{align}
The invariants of the total deformational Finger tensor then become 
\begin{subequations}
\begin{align}
I_1 &= \lambda_0^2(\text{D}\lambda^2 + 3-\text{D}), \label{eq:invi1-d}\\
I_2 &= \lambda_0^4\left[\frac{\text{D}^2-\text{D}}{2} \lambda^4 +\text{D}(3-\text{D})\lambda^2+\frac{(\text{D}-3)(\text{D}-2)}{2}\right], \label{eq:invi2-d}\\
I_3 &= \lambda_0^6\lambda^{2\text{D}}. \label{eq:invi3-d}
\end{align}
\end{subequations}
We plotted the stresses with respect to the stretch ratio $\lambda = R/R_\text{ps}$ in Fig.~\ref{fig:EScompressible}bdf with a pre-stretch $\lambda_0 = 1.2$. The pre-stretched network behaves similarly to the dry network for all models. Pre-stretching would become important in the next section, where a swollen gel ($\lambda_0 = \lambda_\text{sw}$, which is determined by the swelling equilibrium) undergoes phase separation (so that one phase is further swollen and one phase is shrunk), and we observed here that the degree of pre-stretching does not affect the elastic behavior of the models, except in determining the maximum and minimum stretch ratios for the finite extensible models. 

\section{Polymer Gels} 
\subsection{Free Energy}
Introducing solvent into the network makes the network compressible while sustaining the incompressibility of the entire system. The presence of solvent requires mixing and interaction energies, given by the Flory-Huggins model \cite{flory, doisoftmatterphys}, defined in the current frame by the free energy density
\begin{equation} \label{eq:fhmix}
f_{\text{mix}}= \frac{k_B T}{\text{v}_0}\left[(1-\phi)\ln(1-\phi) + \chi \phi (1-\phi)\right], 
\end{equation}
where $\phi$ is the polymer volume fraction and $\text{v}_0$ is the molecular volume of the solvent, which we assume is the same volume as that of the Kuhn monomer. Here we omit the $(1/N)\phi \ln \phi$ term in the classic Flory-Huggins model because instead of a polymer solution, all of the polymer strands are crosslinked into a single network, and its translational energy is replaced by the elastic energy \cite{Dimitriyev_2019}. In several other studies of LLPS with elasticity, this translational term of the polymer was actually (erroneously) included, in order to ensure the existence of the common tangent construction of the free energy curve \cite{D2SM01101H, PhysRevX.14.021009,PhysRevLett.125.268001}. In the simplest case \cite{flory} the dimensionless parameter $\chi$ accounts for the interaction energies between different pairs of species in the system:
\begin{equation}
\chi(T) \equiv -\frac{z}{2k_B T} \Delta \epsilon.
\end{equation}
Here, $z$ is the number of neighboring particles in a lattice model, and $\Delta \epsilon \equiv \epsilon_{pp} + \epsilon_{ss} - 2\epsilon_{ps}$ is the average total energy difference between polymer-polymer, solvent-solvent, and polymer-solvent molecules.
We use a particularly simple dependence of $\chi$ on temperature; in general one expects the form $\chi=A-B/T$ where $A,B$ respectively depend on the entropy and enthalpy of interactions, and detailed comparisons with phase diagrams show that a better phenomenological description includes a concentration-dependent $\chi(T,\phi)$ \cite{TF9716702275, 10.1063/1.460672, ermanFlory1986}. 

To more easily discuss phase separation in polymer gels, the free energy density should be shown in the current (deformed) frame $dV$, because the equilibrium condition of phase separation requires the true stress and the chemical potential as defined in the current frame. We can convert the elastic free energy density, thus far defined in the initial reference frame $dV_0$, into the current space using the relation given by Eq.~\eqref{eq:fref-fcur}, $dV = dV_0\sqrt{I_3}$. The invariant $I_3$ is related to the polymer concentration by:
\begin{equation} \label{eq:phi0}
\frac{\phi_\text{id}}{\phi} = \sqrt{I_3},
\end{equation}
 where $\phi_\text{id}$ is the volume fraction of the polymer network consisting of ideal chains. We have considered elastic models in which the dry (unswollen) state is unstressed for $\lambda=1$, which means that $\phi_\text{id} = 1$. Note that this is not always the case: if the polymer is fabricated in a $\theta$-solvent and then washed and dried, the polymer strand configuration in the dry polymer network would not follow the Gaussian distribution, \textit{i.e.} the polymer strand end-to-end distance $\langle R^2\rangle \neq Na^2$, so that $\lambda = 1$ at $\phi < 1$. 
Depending on how the network is prepared (\textit{e.g.} in solution or in melt state), the polymer chains may be prestrained \cite{LinOlsen2019,doi:10.1021/acs.macromol.8b01676,doi:10.1021/acsmacrolett.0c00909} before swelling and phase separating. In this paper, we only consider an initial state that is consistent with our elastic models, which define the deformation to be a deviation from the dry ideal chain state where $\langle R_0^2\rangle = Na^2$. Now we can write the elastic free energies $f^0_\text{el}$ as
\begin{equation}
    \mathcal{F}_\text{el}=\int dV f_\text{el} = \int dV_0 f_\text{el}^0 = \int dV \frac{\phi}{\phi_{\textrm{id}}}f_\text{el}^0 = \int dV \phi f_\text{el}^0,
\end{equation}
and the total free energy density as the sum of the elastic free energy $f_{\text{el}}$ and the mixing energy $f_\text{mix}$, 
\begin{align}
    f&= f_{\textrm{el}}+f_{\textrm{mix}}, \label{eq:fffmix}\\
    &=\phi f^{0}_{\textrm{cf}}+\frac{\Phi}{2}n_{x0}k_BT\alpha\phi\ln\phi+f_\text{mix}. \label{eq:fffmixB}
\end{align}

\subsection{Stress}
The stress tensor after a strain $\varepsilon_{\alpha\beta}$ that includes a volume change is found most easily by first considering the free energy in the reference configuration:
\begin{align}
\begin{split} \label{eq:tstresscomp_a}
\delta\mathcal{F}\Big|_{\varepsilon_{\alpha\beta}} &= \delta \int \left(\sqrt{I_3}dV_0\right) f =\int dV_0 \: \delta\left( \sqrt{I_3}f\right),\\
&= \int dV \left(f\delta_{\alpha \beta}+\frac{\partial f}{\partial \varepsilon_{\alpha \beta}} \right) \varepsilon_{\alpha \beta},
\end{split}
\end{align}
where we have used $\delta I_3 = 2 \text{Tr}\varepsilon$. Following
Eq.~\eqref{eq:tstressPA}, the stress tensor components for principal extensions $\lambda_i$ are given by
\begin{align}
\sigma_{ii} &= f\delta_{ii}+\lambda_i \left.\frac{\partial f}{\partial \lambda_{i}}\right|_{\lambda_{j\neq i}},\label{eq:tstresscomp_diag} 
\end{align}
where we do not sum over the repeated indices $i$ in the second term above (see Supplementary Information Sec.~1.2 for further explanation).

To find the stress in terms of composition derivatives, we first note that $\phi=\left(\prod_j\lambda_j\right)^{-1}$. From this we can derive
\begin{equation}
\left.\frac{\partial\phi}{\partial\lambda_j}\right|_{\lambda_{k\neq j}} = -\frac{\phi}{\lambda_j}.
\end{equation}
Hence, the stresses in the unconstrained directions $j$ can be written entirely in terms of $\phi$:
\begin{align}
\sigma_{jj} &= f-\phi \frac{\partial f}{\partial \phi}, \label{eq:tstresscomp_phi}
\end{align}
which is simply the negative osmotic pressure $-\Pi_{\textrm{osm}}$ if the free energy is the classic Flory-Huggins mixing energy of a polymer-solvent solution. The stress in the constrained dimension will change due to the change in polymer concentration:
\begin{equation} \label{eq:tstresscomp_uc}
\sigma_{ii}(\lambda_i = 1) = 2\phi (f^0)', 
\end{equation}
where $(f^0)'$ is a function that consists of the first derivatives of the free energy in the initial frame with respect to the invariants (see the Supplementary Information Sec.~1.2 for a detailed derivation):
\begin{equation} \label{eq:df0_uc}
 (f^0)' = \frac{\partial f^0}{\partial I_1} + \frac{\partial f^0}{\partial I_2}(I_1 - 1) + \frac{\partial f^0}{\partial I_3} I_3. 
\end{equation}
Even though the stress(es) in the constrained direction(s) can still change with deformation, because we impose some external forces on the gel to hold it undeformed in this/these direction(s), they are of less interest to us than the stress(es) in the unconstrained directions. For this reason, in this study we will focus on the stress in the constrained direction(s). If there is no elasticity, so the free energy density is only a function of composition $\phi\sim1/\sqrt{I_3}$, one can show that the stress tensor recovers the form of the osmotic pressure $\sigma_{ii}=f-\phi\partial f\partial\phi = -\Pi_\text{osm}$.

We list the stresses for the D-dimension deformation below:
\begin{subequations} 
\begin{align}
\sigma_{xx} &= f-\phi \frac{\partial f}{\partial \phi},\quad \sigma_{yy}=\sigma_{zz}=2\phi (f^0)'&&(1\text{D}) \label{eq:sigxx1d}\\
\sigma_{xx} &= \sigma_{yy} =f-\phi \frac{\partial f}{\partial \phi}, \quad\sigma_{zz}=2\phi (f^0)'&&(2\text{D}) \label{eq:sigxxyy2d}\\
\sigma_{xx} &= \sigma_{yy}=\sigma_{zz} = f-\phi \frac{\partial f}{\partial \phi}. &&(3\text{D}) \label{eq:sigxxyyzz3d}
\end{align}
\end{subequations}

To compare models, we will use the parameters listed in Table.~\ref{tab:fig-parameters}, in which all models have the same Young's modulus when the material is `dry', \textit{i.e.} $\phi_\text{id} = 1$, and the network is assumed to obey Gaussian statistics in this dry state. The dry polymer network is then placed in a solvent bath, upon which it swells isotropically to network volume fraction $\phi_{\text{sw}}<1$ with the corresponding swelling ratio $\lambda_{\textrm{sw}}=\phi_{\textrm{sw}}^{-1/3}$. For subsequent swelling, the elastic free energy should be evaluated using the total deformation gradient $\boldsymbol{F}_{\textrm{tot}}=\boldsymbol{F}\boldsymbol{F}_{\textrm{sw}}$, where $\boldsymbol{F}$ is the subsequent deformation after swelling. The Finger tensor is $\boldsymbol{W}_\text{tot} = \boldsymbol{F}_{\textrm{tot}} \boldsymbol{F}_{\textrm{tot}}^T = \boldsymbol{F} \mathbf{W}_\text{sw}\boldsymbol{F}^T$. 
This leads to 
\begin{subequations}
\begin{align}
I_1(\boldsymbol{W}_{\textrm{tot}})&= \lambda_{\textrm{sw}}^{2}I_1(\lambda)\\
I_2(\boldsymbol{W}_{\textrm{tot}})&= \lambda_{\textrm{sw}}^{4}I_2(\lambda)\\
I_3(\boldsymbol{W}_{\textrm{tot}})&= \lambda_{\textrm{sw}}^{6}I_3(\lambda).
\end{align}
\end{subequations}

The swollen network-solvent gel can then be taken out of the solvent bath to impose constrained swelling. 
In this case, for D $=1$ we have $\boldsymbol{F}_{\textrm{tot}}=\lambda_{\textrm{sw}}\,\textrm{diag}(\lambda,1,1)$,  for D $=2$ we have $\boldsymbol{F}_{\textrm{tot}}=\lambda_{\textrm{sw}}\,\textrm{diag}(\lambda,\lambda,1)$, etc.

The volume fraction $\phi$ of  polymer and invariants of the deformation gradient obey:
\begin{subequations}\label{eq:invariants-constrained-phi}
\begin{align}
\frac{1}{\phi} &= \sqrt{I_3} =\lambda_{\textrm{sw}}^{3} \lambda^{\text{D}} = \frac{1}{\phi_{\text{sw}}} \lambda^{\text{D}}\label{eq:phirel3}\\
\begin{split} \label{eq:phirel1}
I_1 &= \lambda_{\textrm{sw}}^2 \left(\text{D}\lambda^2+3-\text{D}\right) \\
&= \phi_{\text{sw}}^{-\frac{2}{3}} \left[ \text{D}\left(\frac{\phi_{\text{sw}}}{\phi}\right)^{\frac{2}{\text{D}}} + 3 - \text{D}\right]
\end{split}\\
\begin{split} \label{eq:phirel2}
I_2 &= \frac{\phi_{\text{sw}}^{-\frac{4}{3}}}{2}\Big[\text{D}(\text{D}- 1)\left(\frac{\phi_{\text{sw}}}{\phi}\right)^{\frac{4}{\text{D}}} +2\text{D}(3-\text{D})\left(\frac{\phi_{\text{sw}}}{\phi}\right)^{\frac{2}{\text{D}}}\\
& \qquad \;\;+(2 - \text{D} )(3 - \text{D} )\Big]. 
\end{split}
\end{align}
\end{subequations}

\subsection{Equilibrium conditions}

\subsubsection{Swelling equilibrium} \label{subsec:prepro}
The swelling equilibrium concentration $\phi_{\text{sw}}$ at a given temperatures $T_0$ satisfies the stress free condition inside the network and stress balance (with the external osmotic pressure) at the surface:
\begin{align}
&\left.\frac{\partial \sigma_{ii}}{\partial R_i}\right\vert_{\text{in gel}} = 0, \label{eq:mecheqbA}\\
&\left.\bm{\sigma}\right|_{\textrm{surface}} \cdot \bm{\hat{n}} = \bm{0}, \label{eq:mecheqbB}
\end{align}
where $\bm{\hat{n}}$ is the normal vector of the material surface; and we assume that the osmotic pressure of the solvent bath is zero. In a homogeneous gel, Eq.~\eqref{eq:mecheqbA} is always satisfied so that Eq.~\eqref{eq:mecheqbB}  determines $\chi_0(T_0)$:
\begin{equation} 
\begin{split} \label{eq:chi0}
    \chi_0(T_0) = \frac{1}{\phi_{\text{sw}}^2}\left[\frac{\text{v}_0}{k_B T }\left(-f_{\text{el}}(\phi_{\text{sw}})+\phi_{\text{sw}} \left.\frac{\partial f_{\text{el}}}{\partial \phi}\right\vert_{\phi=\phi_{\text{sw}}}\right)\right. \\\quad
    - \ln(1-\phi_{\text{sw}})-\phi_{\text{sw}}\Bigg].
\end{split}
\end{equation}
where the gel swells isotropically (D $=3$) from the dry state $\phi = 1$ to a swollen state $\phi_\text{sw} < 1$. Once the swelling equilibrium has been attained at a given temperature $T_0$, we can determine $\chi(T)$ at another temperature $T$  via
\begin{equation} \label{eq:chi}
    \chi(T) = \chi_0(T_0)\frac{T_0}{T}. 
\end{equation}

The swelling equilibrium depends on the crosslink density and temperature $n_{x0}\kbT$, which for a given model can be determined from Young's modulus or the shear modulus of the dry gel; the $\chi$ parameter (itself a function of temperature and, in principle, concentration); the osmotic contribution to elasticity parameterized by $\alpha$ in Eq.~\eqref{eq:fingel}; and the specific elastic model. The parameters \{$n_{x0}k_B T $, $\alpha$, $\chi_0$\} are needed when we compare the free energy densities and the stresses of the six models, and to generate phase diagrams. Since $n_{x0} k_B T$ is related to Young's modulus $E$ by Eq.~\eqref{eq:nx0nums}, which can be independently measured in tensile tests, we will treat $E$ as a known parameter. 
 
If gel swelling experiments give us the swelling concentration $\phi_\text{sw}$, and we assume that the prefactor $\alpha$ is known, we can find the value of $\chi_0$ at a given $T_0$ using Eq.~\eqref{eq:chi0}. The problem with this approach is that 1) the value of $\alpha$ is still widely debated, and 2) $\chi_0$ will differ for different models, which is not physical since $\chi$ does not depend on the structural properties of the network, only the inter-molecular chemistry and temperature. However, we note that $n_{x0} \text{v}_0 \sim \text{v}_0/\xi^3\ll1$, where $\xi$ is the mesh size of the gel, so that the contribution from the elastic free energy is generally much smaller than that from the mixing energy, until the deformation becomes extreme. As a result, $\chi_0$ is predominantly determined by the mixing entropy, so that the difference in the inferred $\chi_0$ between different elastic models would be negligible (see second to last row in Table.~\ref{tab:fig-parameters}).

Another possibility is that, for a given $E$, a known $\chi_0(T_0)$ and an agreed-upon value for $\alpha$, we can find $\phi_\text{sw}$ with Eq.~\eqref{eq:chi0}. The last line in Table.~\ref{tab:fig-parameters} lists the $\phi_\text{sw}$ values given by $E = 40$ kPa, $\alpha = 0.5$ and $\chi_0(320K) = 0.869$. With this approach, the problem with $\alpha$ being chosen arbitrarily remains, but the differences in $\phi_\text{sw}$ are still small, so it is practically not very different from the first approach with the varying $\chi_0$.

\begin{figure*}[htb]
\centering
\includegraphics[width=0.9\textwidth]{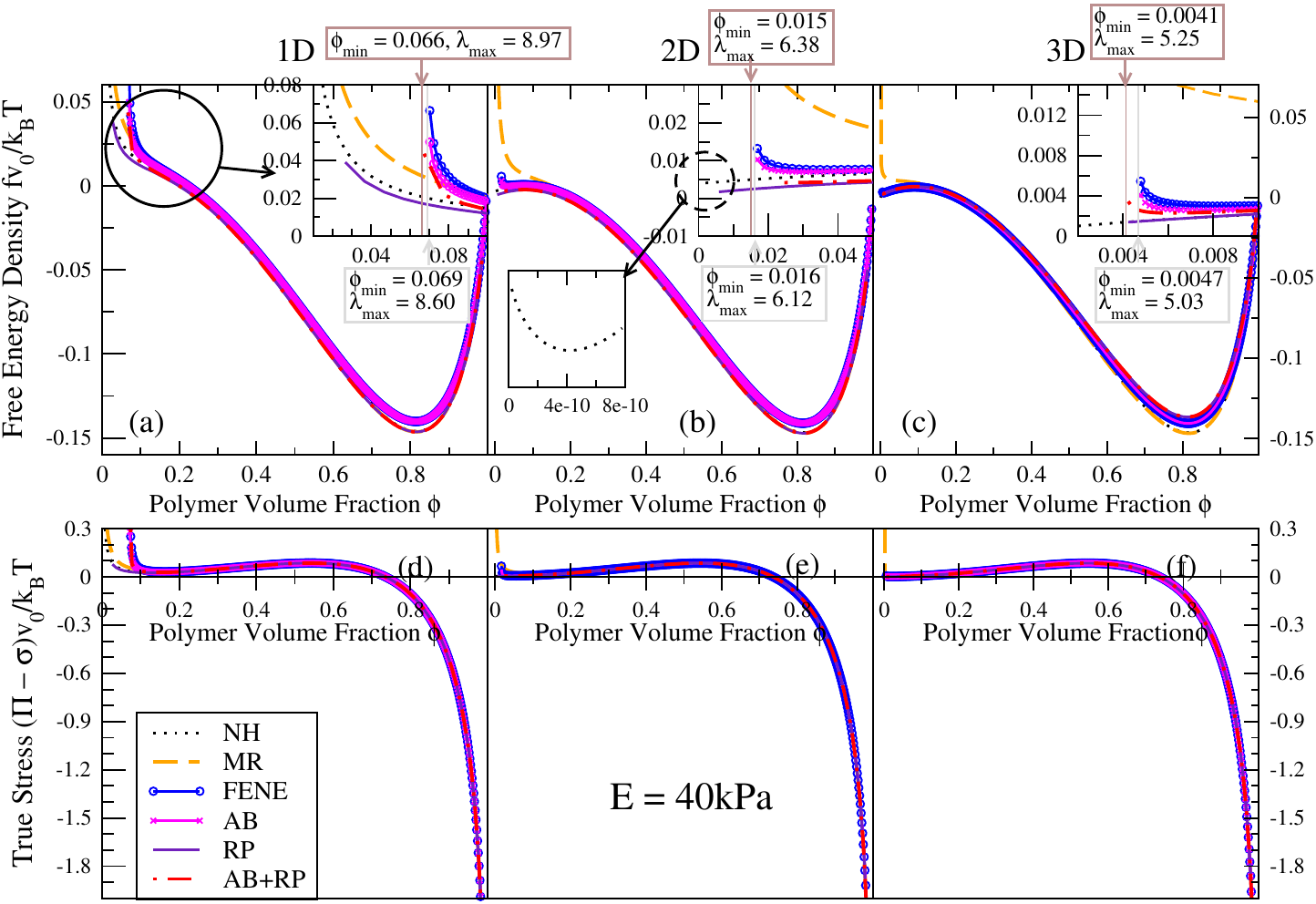}
\caption{Top: Free energy densities of PDMS-Solvent mixture in the current frame. Bottom:  true stress in the direction(s) of the deformation. All models have a Young's modulus of  40 \textrm{kPa}, a mixing parameter $\chi = 1.1$, and initial network volume fraction $\phi_{\text{sw}} = 0.595$. $\alpha$ values are listed in Table~\ref{tab:fig-parameters}. The grey (NH, FENE, AB) and brown (RP, AR+RP) vertical lines indicate the minimum $\phi_\text{min}$ (or maximum $\lambda_\text{max}$).}\label{fig:ESphi}
\end{figure*}

The third approach is to set the values for $\phi_\text{sw}$ and $\chi_0(T_0)$, and then find $\alpha$ values for different elastic models for a given $E$. Since our goal in this study is to compare elastic models to determine which model is the best to use for the network-solvent phase separation problem, this is the approach we take for the rest of the paper. We first identify a specific reference point $\{E, \phi_{\text{sw}}, T_0\}$ and a $\chi_0$ value that result in $\alpha_\text{NH} = 0.5$ for the NH model and $\alpha$ values listed in Table~\ref{tab:fig-parameters} for the other elastic models. The $\alpha$ values are calculated to satisfy the stress free condition given by Eq.~\eqref{eq:chi0} at the boundaries in contact with the solvent bath. Only  the MR model has a value for $\alpha$ that departs significantly from the NH model ($\alpha = 0.5$). This is because the MR model contains two terms $C_1$ and $C_2$ that contribute to the Young's modulus, yielding a smaller density of crosslinks $n_{x0}$ (Eq.~\eqref{eq:MRparameter} and Table~\ref{tab:fig-parameters}, line 1), which then necessitates a larger $\alpha$ in order to obtain the same osmotic effect. We emphasize that the MR model has limited applicability, since it was only designed to fit data on the uniaxial stretch of incompressible networks.


\subsubsection{`Naive' phase coexistence within the gel}
To construct phase diagrams for models that support phase separation, we need to apply the equilibrium conditions for the coexisting phases. When a sharp interface exists between two homogeneous phases, coexistence is given by the common tangent construction:
\begin{align}
    -f(\phi_1) +\phi_1 \left.\frac{\partial f(\phi)}{\partial \phi}\right \vert_{\phi_1} &= -f(\phi_2) +\phi_2 \left.\frac{\partial f(\phi)}{\partial \phi}\right \vert_{\phi_2}, \label{eq:prseq}\\
    \left.\frac{\partial f(\phi)}{\partial \phi}\right \vert_{\phi_1}&=\left.\frac{\partial f(\phi)}{\partial \phi}\right \vert_{\phi_2}, \label{eq:cpeq}
\end{align}
where Eq.~\eqref{eq:prseq} is the pressure equilibrium and Eq.~\eqref{eq:cpeq} is the chemical potential equilibrium. 


Composition changes $\delta\phi=\phi-\phi_0$ in a polymer gel must be accompanied by an elastic deformation, according to $\phi_\text{sw}/\phi=\lambda^\text{D}$, or $\delta\phi=-\phi\,\textrm{Tr}\boldsymbol{\varepsilon}$ for small deformation. The elastic energy cost is a long range interaction without an intrinsic length scale, which implies that without special treatment of the material geometries that would confine the length scale of the interfacial width between the phase separated domains, there would not a distinctive sharp interface between the domains. In the absence of a sharp interface, the common tangent method cannot be applied to determine the phase coexistence conditions. Nevertheless, we will use the naive common tangent construction as a starting point to understand phase separation. A more precise calculation should incorporate the elastic deformation around the coexisting domains, analogous to the well-known Gent calculation \cite{doi:10.1098/rspa.1959.0016,Horgan:2004aa, HORGAN1992279, Ronceray_2022, KOTHARI2020104153, Zhu:2018aa, D2SM01101H, PhysRevE.107.024418}. This  leads to nonuniversal behavior that depends on the sizes and geometries of the coexisting phases and the sample. 

\subsection{Naive constrained-dimension phase coexistence} \label{subsec:fe&str}
We now study the predictions of `naive' phase coexistence for different models, in different constrained-dimension geometries: a 1D phase separation whereby a thin sheet changes its thickness, a 2D phase separation in which a thin rod changes the rod thickness (swelling in the radial direction), and a 3D phase separation in which two coexisting regions swell/shrink isotropically. 

Fig.~\ref{fig:ESphi} shows the total free energy densities $f$ in the current frame and stresses in the unconstrained direction(s), as a function of concentration, for 1D, 2D, and 3D constrained deformations. Calculations are performed for all six models, with a PDMS network with Young's modulus $E = 40$ kPa, $\chi_0 = 0.8692$ at $T_0 = 320K$, $\phi_\text{sw} = 0.595$, and $\alpha=\alpha^{\ast}$ as listed in the third to last line of Table~\ref{tab:fig-parameters}. Free energy density curves with the classic `W' shape (possessing a concentration range with negative curvature $\partial^2f/\partial\phi^2<0$) permits the use of the common tangent construction to determine coexisting compositions.

For a 1D deformation, common tangents are possible for all six models. The free energy is much higher at low polymer concentration for the strain stiffening models (Mooney-Rivlin, FENE, Arruda-Boyce, and AB+RP) than for the Neo-Hookean (NH) model, because of the finite extensibility of the crosslink strands. Hence, phase coexistence in the strain stiffening models would have a narrower range of coexisting concentrations when the binodal extends to very dilute network compositions. However, it is worth noting that if we draw a common tangent line for the NH model, the low concentration $\phi_1$ in co-existence would be smaller than the theoretical minimum $\phi_{\textrm{min}}\sim1/\lambda_{\textrm{max}}$ that the network can support, where $\lambda_{\textrm{max}}\sim1\sqrt{N}$ is the maximum deformation that a single strand can sustain (see Appendix~\ref{appx:maxl}). In other words, the network strands would technically break before reaching the dilute coexisting concentration $\phi_1$ calculated from the NH model for a 1D deformation. The brown arrows in the inset of Fig.~\ref{fig:ESphi}abc show the composition corresponding to breakage. We estimate $\lambda_{\textrm{max}}$ for the NH model by calculating the strand length $N$ from the elastic modulus,  Eq.~\ref{eq:nreNH}.
 
In a 2D deformation, the common tangent conditions cannot be met for the slip tube model, because the curvature of its free energy density only changes sign once. In a 3-D deformation only the finite extensibility (FENE, AB, AB+RP) models and the MR model (from the higher order elastic term controlled by $C_2$) support the common tangent construction. 

The NH and RP models do not support a common tangent construction in 3D. Careful inspection of the free energy densities shows that the negative curvature persists even for $\phi=0$ in the dilute limit of these two models (the two models with Gaussian chains). This is due to the network's volume change (\textit{i.e.} the $1/\sqrt{I_3}$ prefactor preceding the elastic free energy), which modifies the free energy density in the current frame. For the NH model, the elastic free energy in the current frame (see Eqs.~\ref{eq:fref-fcur}, \ref{eq:n-h-g}, \ref{eq:fingel}, \ref{eq:invariants-constrained-phi}) is (omitting unimportant prefactors)
\begin{align}
    \frac{f_{\textrm{NH}}}{n_{x0}\kbT} &\sim \frac{1}{\sqrt{I_3}} (I_1-3-\alpha\ln I_3),\\
    &= \text{D}\phi^{(1-2/\text{D})} -\text{D}\phi +2\alpha\phi\ln\phi, \\
    &= 
\begin{cases}
    \dfrac{1}{\phi} - \phi + 2\alpha \phi\ln\phi & (\text{D}=1), \\[6truept]
    -2\phi + 2\alpha \phi\ln\phi & (\text{D}=2), \\[6truept]
    3\phi^{1/3} - 3 \phi + 2\alpha \phi\ln\phi &(\text{D}=3),
\end{cases}
\end{align}
to which we add the mixing free energy (Eq.~\ref{eq:fhmix}) to calculate the curvature of the total free energy, $\partial^2f_{\textrm{NH}}/\partial\phi^2$. If the curvature is negative as $\phi\rightarrow0$ then the common tangent construction is not possible. The curvatures are given by
\begin{equation}
    \label{eq:f-curve}
    \frac{1}{n_{x0}\kbT}\frac{\partial^2 f_{\textrm{NH}}}{\partial\phi^2} \sim 
    \begin{cases}
    \dfrac{9+2\alpha\phi}{\phi^3} + \dfrac{1}{n_{x0}\text{v}_0}\left[\dfrac{1}{1-\phi}-2\chi\right] & (D=1), \\[12truept]
    \dfrac{2\alpha}{\phi} + \dfrac{1}{n_{x0}\text{v}_0}\left[\dfrac{1}{1-\phi}-2\chi\right], & (D=2), \\[12truept]
    -\dfrac{2(1+\alpha\phi^{2/3})}{3\phi^{5/3}} + \dfrac{1}{n_{x0}\text{v}_0}\left[\dfrac{1}{1-\phi}-2\chi\right] & (D=3).
     \end{cases}
\end{equation}

Phase coexistence is possible in 1D and 2D for the NH model, but not in 3D, since the curvature of the free energy is always negative in the limit $\phi\rightarrow0$. In 2D, the dilute phase has very low polymer concentrations $\phi_{\textrm{dil}}\sim\alpha(n_{x0}\text{v}_0)/(2\chi-1)\sim\alpha (a/\xi)^3/(2\chi-1)\ll1$. These low volume fractions correspond to molecular deformations that exceed the limit of validity of the Gaussian approximation that leads to the Neo-Hookean model (see arrows in Fig.~\ref{fig:ESphi}ab).

In Fig.~\ref{fig:ESphi}, the free energy densities and the stresses of the six models almost superpose, except at low polymer concentrations. This is because the elastic energy contributes much less than the mixing energy except at very low polymer concentrations where the polymer incurs significant stretching. For each model, the free energy density and the stress decrease as the deformation dimension increases, which agree with the result of the isolated elastic free energy density shown in Fig.~\ref{fig:EScompressible}. The competition between the increase in the elastic energy and the increase in volume also explains why the free energy density decreases (especially in the low polymer concentration region) as the deformation dimension increases. As a result, it costs less energy to phase separate in a higher dimension deformation, which is easier to see with phase diagrams that we will present in the next section. 

These calculations suggest that strain-stiffening due to finite extensibility is necessary to stabilize phase separation in a 3D (isotropic) deformation during phase separation. In principle, one can also stabilize phase separation by choosing a suitable concentration dependence of the $\chi$ parameter \cite{10.1063/1.460672}. A recent study \cite{Carla_unpublished_work} included a concentration-dependent $\chi$ parameter in modeling the phase behavior of an uncrosslinked polymer solution, but even with such a dependence, a strain-stiffening elastic model was required to describe the phase separation seen in a cross-linked version of the same mixture. Hence, it is likely that finite elasticity is required in order to generically stabilize phase separation sufficiently far below the critical point at which the dilute phase is forced to undergo significant stretching. This is explored further in Appendix~\ref{appx:varchi}. 

\begin{figure}[htb]
\centering
\includegraphics[width=0.45\textwidth]{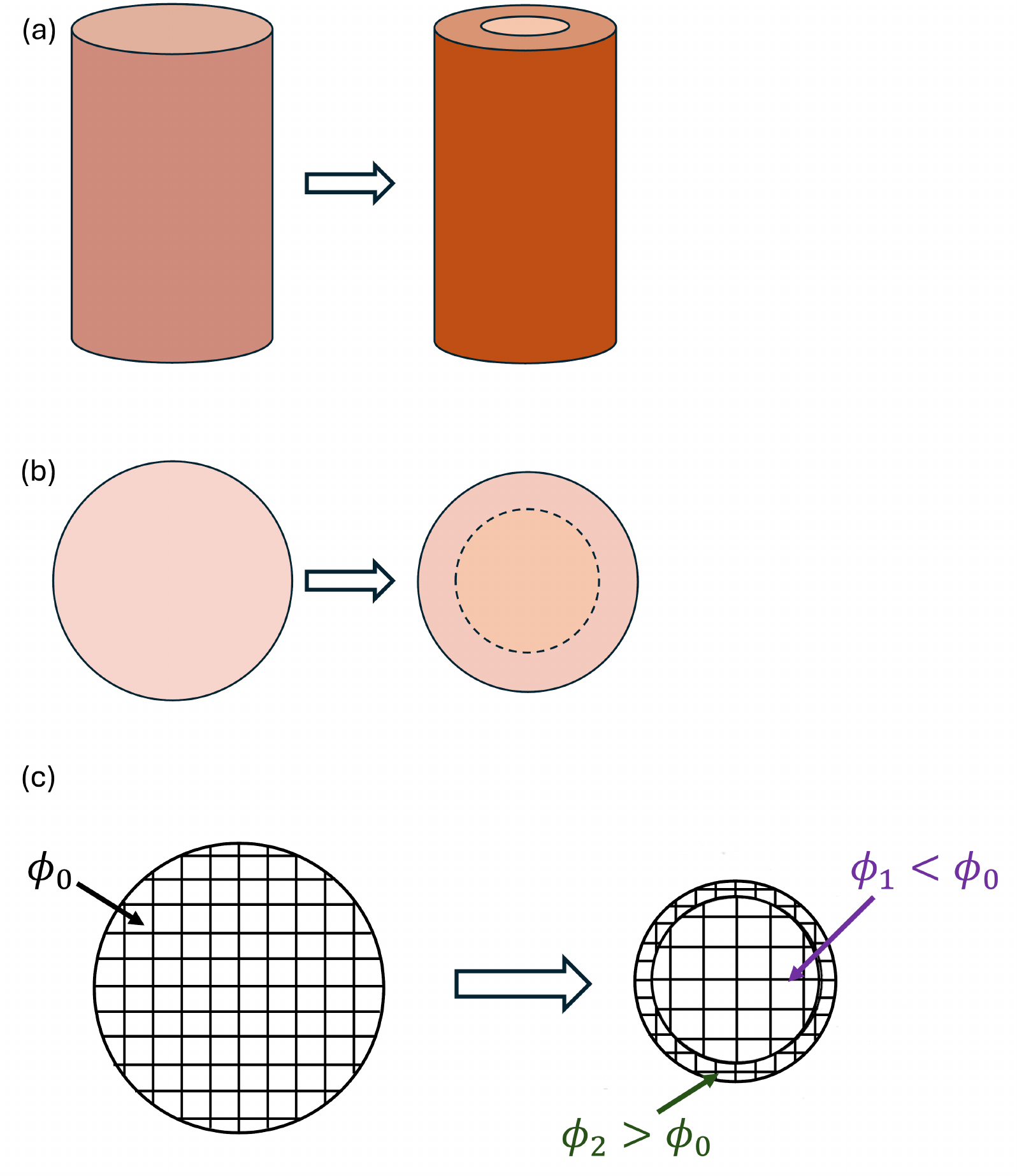}
\caption{Polar deformations that satisfy Eq.~\eqref{eq:prseq}: (a) phase separated cylinder, (b) phase separated sphere, and (c) cross section schematics of the polar phase separation, the polymer strands are disconnected at the interface.}\label{fig:real23dps}
\end{figure}

\subsection{Polar deformation geometry}

When we take a closer look at the deformation geometry proposed in Fig.~\ref{fig:cddsc}, we find that the stress condition obtained in Eq.~\eqref{eq:prseq} balances the stresses in the unconstrained direction of the two phases, so that the stresses in the perpendicular direction for the thin sheet and the radial directions for the rod are equated, instead of the stresses perpendicular to and across the interfaces. Therefore, Fig.~\ref{fig:cddsc} does not adequately represent the phase separation described by Eqs.~(\ref{eq:cpeq}, \ref{eq:prseq}), even if we ignore the issue of lack of sharp interfaces, which leads us to reconsider how the two phases are in contact with each other. If we stack 1D deformed thin sheets like lasagna, with alternating $\phi_1$ and $\phi_2$ layers, then the stress balance at the interface satisfies Eq.~\eqref{eq:prseq}. Similarly, for 2D and 3D deformations we need to nest the two phases `inside' of each other to satisfy the stress balance at the interfaces. Fig.~\ref{fig:real23dps} shows this new configuration for cylinders and spheres. We can construct this configuration from two identical infinitely long cylindrical gel rods with the same initial concentration $\phi_0$. One gel rod is swollen isotropically to a new state $\phi_1 < \phi_0$, and the other is shrunk to $\phi_2 > \phi_0 $. We then hollow out the shrunken rod with polymer concentration $\phi_2$ and replace the hole by inserting the one with concentration $\phi_1$. The resulting system fits our definition of a 2D constrained-deformation deformation. Similarly, a 3D deformation can be realized by inserting a sphere inside another (Fig.~\ref{fig:real23dps}b). 

We acknowledge that even though the new polar geometry satisfies the stress balance condition, the issue with the lack of a sharp interface remains, and we have also allowed an nonphysical configuration that does not respect continuity of the strain field between the two phases. As we can see in Fig.~\ref{fig:real23dps}c, there is a discontinuity of the polymer strands at the interface, which would require some surface energy cost to break them. Because the polymer strands in the shell cannot attach to those in the center in a one-to-one fashion (since there are more strands outside than inside), there must be some surface treatment to make the strands somehow attach to the interface. Despite that, we suggest that the interfacial free energy caused by such surface treatment is negligible compared to the bulk free energy of the (semi-)infinite material, and this polar phase separation geometry is a good enough representation of the 2D and 3D constrained-deformation deformations. We also need to point out that spherical geometry is often used in calculations concerning cavity growth inside elastic materials \cite{doi:10.1098/rspa.1959.0016,Horgan:2004aa, HORGAN1992279, Ronceray_2022, KOTHARI2020104153}, which is probably a better interpretation of the gel phase separation studied by the Dufresne group \cite{Fernandez-Rico:2024aa,Rosowski:2020ab,D0SM00628A,doi:10.1126/sciadv.aaz0418,PhysRevX.8.011028,Style:2015aa,Jensen14490}. 

\begin{figure*}
\centering
\includegraphics[width=0.9\textwidth]{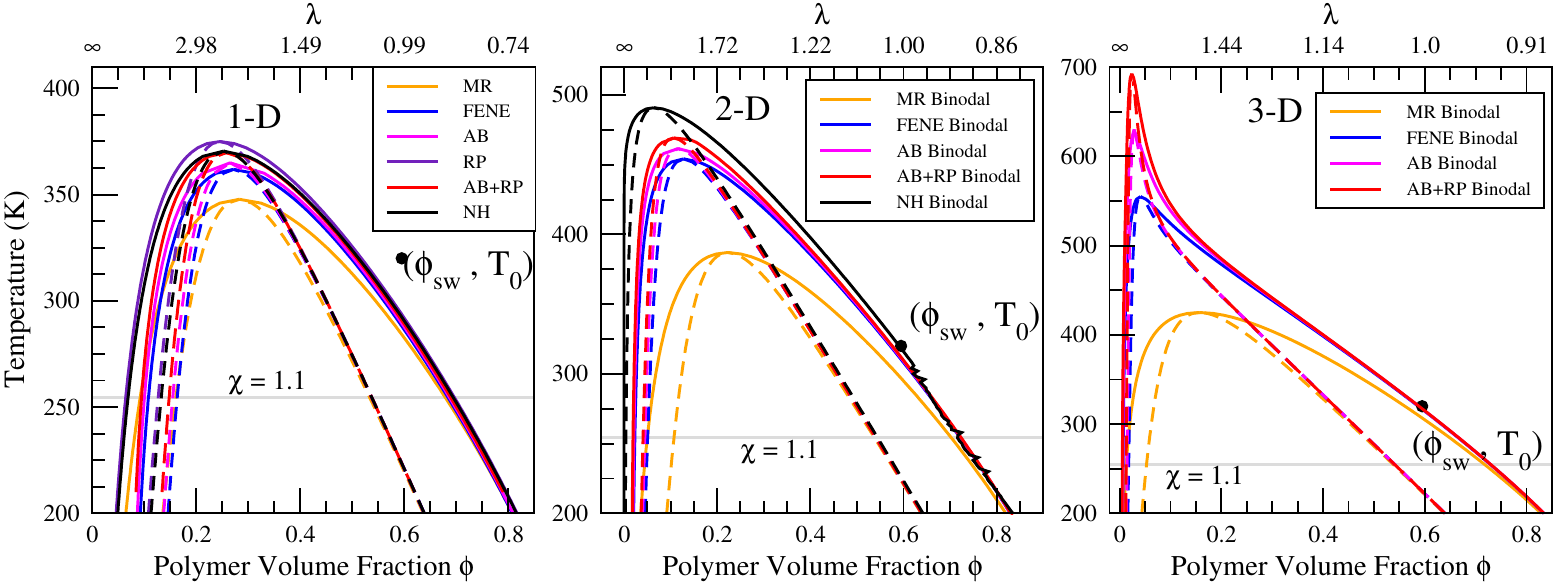}
\caption{Phase Diagrams of different models undergoing 1D (left), 2D (middle) and 3D (right) deformations, for a $E = 40$ kPa PDMS (parameters listed in Table.~\ref{tab:fig-parameters}) with the same $\chi_0$ and $\phi_\text{sw}$ at $T_0$ and varying $\alpha=\alpha^{\ast}$ . The gray line indicates the value of $\chi = 1.1$ used in plotting the free energies and stresses in Fig.~\ref{fig:ESphi}.}\label{fig:pd3dim}
\end{figure*}

\section{Phase Separation }\label{sec:pd}
\subsection{Phase Diagram}

To calculate the phase diagrams in different deformation dimensions (if possible) we use the same procedure used above to plot the free energies and stresses in Fig.~\ref{fig:ESphi}. We assume that an $E = 40$ kPa PDMS network, whose ideal state is the dry state $\phi_\text{id} = 1$, will swell in some solvent to have a swelling concentration $\phi_\text{sw} = 0.595$ and $\chi_0 = 0.8692$ at $T_0 = 320 K$, and then we determine $\alpha=\alpha^{\ast}$ so that the swelling equilibrium Eq.~\eqref{eq:chi0} is satisfied for each model (Table~\ref{tab:fig-parameters}). The parameters $\{E, \phi_\text{sw}, \chi_0, \alpha\}$ are then used to produce the phase diagrams in Fig.~\ref{fig:pd3dim}.

Fig.~\ref{fig:pd3dim} shows phase diagrams of the three constrained-dimension deformations for all models that permit (naive) phase coexistence. In general, `softer' networks (NH, slip-tube) phase separate more easily: larger unstable (enclosed by binodal curves) and metastable (enclosed between binodals and spinodals) regions allow phase separation to occur at higher temperature. This is due to the smaller elastic energy costs to deform Gaussian chains compared to deforming freely jointed chains to the same degree. When comparing the same model across deformation dimensions, we find that as the deformation dimension increases, the critical points move lower in polymer concentration and higher in temperature. The decrease in the critical polymer concentration with increasing dimension is mainly the result of the necessity to reach a similar extension ratio $\lambda_c$ in the unconstrained direction(s). We find that $\lambda_c$ is roughly the same in all deformation dimensions, but because a higher D leads to a larger change in concentration, according to $\phi= \phi_{\text{sw}} \lambda^{-\text{D}}$, the left shift (to a lower critical concentration $\phi_c$) is much more prominent in the phase diagrams in Fig.~\ref{fig:pd3dim}. For example, the critical concentrations of the AB model in D-dimension are $\phi_c^{1\text{D}} = 0.265$, $\phi_c^{2\text{D}} = 0.116$, $\phi_c^{3\text{D}} = 0.029$, but the corresponding stretch ratios are very similar: $\lambda^{1\text{D}}_c = 2.24$, $\lambda^{2\text{D}}_c = 2.26$, $\lambda^{3\text{D}}_c = 2.75$. The shift in the phase diagram can also be understood as follows: The pressure on the concentrated polymer side of the interface is similar across all deformation dimensions since it is dominated by the osmotic pressure. To balance the pressure on the polymer rich phase, the lower-dimensional deformations require a larger change in the polymer concentration in the polymer-poor phase to achieve the same elastic stress; thus, the gel phase separates at a lower temperature in lower dimensional deformation. This also explains why the spinodals of all models superpose at the higher polymer volume fraction phase (right leg of the spinodal curves) but differ drastically at the lower polymer concentration (left leg of the spinodal curves).

For 3D deformations, the critical temperatures of the finite extensibility models (FENE, AB, and AB+RP)  increase dramatically compared to their 1D and 2D counterparts, with a narrow meta/unstable region caused by the elastic energy always favoring separated phases. The narrowing near the critical points means that at higher temperature, the concentrations of the coexisting phases are very close to each other, thus making this part of the phase diagrams true to the common tangent method since the interface would be thin.

\subsection{Effects of the osmotic elastic energy}

As mentioned in Section \ref{sec:dimension}, the term $n_{x0}k_B T\alpha \ln I_3$ in 
Eq.~(\ref{eq:fingel}) accounts for the entropy associated with the fluctuations of the crosslinks. As shown above in Eq.~\eqref{eq:fffmixB}, the free energy of the osmotic elastic term, in the current frame, can be written as
\begin{equation}\label{eq:fosm_N}
f_{\textrm{el,osm}}=\frac{\kbT}{\text{v}_0}\frac{\phi}{N_{\textrm{eff}}}\ln\phi, 
\end{equation}
the form resembles the translational entropy of a polymer of size $N_{\textrm{eff}}=1/(2\alpha \text{v}_0 n_{x0})\simeq N^{3/2}/(2\alpha)$. The effective size scales with Young's modulus as $N_{\textrm{eff}}\sim\kbT/E$. The value of $\alpha$ has not been settled in the literature; the phantom network model predicts $\alpha=0$ \cite{doi:10.1098/rsta.1976.0001,PANYUKOV19961}, the affine junction model predicts $\alpha=1/2$ \cite{1e691dd5-cdae-38d3-b92f-1f2dc667da23}, and models with more complex topologies can have values $0<\alpha\leq1/2$ \cite{1e691dd5-cdae-38d3-b92f-1f2dc667da23}. Unfortunately, it is very difficult to firmly establish the value of $\alpha$ experimentally\cite{Dimitriyev_2019}. In view of this uncertainty, we briefly study how the value of $\alpha$ affects the critical concentration and critical temperature of PDMS networks with various stiffness. Since vanishing $\alpha$ corresponds to the translational entropy of a very large polymer, we expect that increasing $\alpha$ increases the entropic cost to phase separate and thus decreases $T_c$. 

\begin{figure}[hbt!]
\centering
\includegraphics[width=0.47\textwidth]{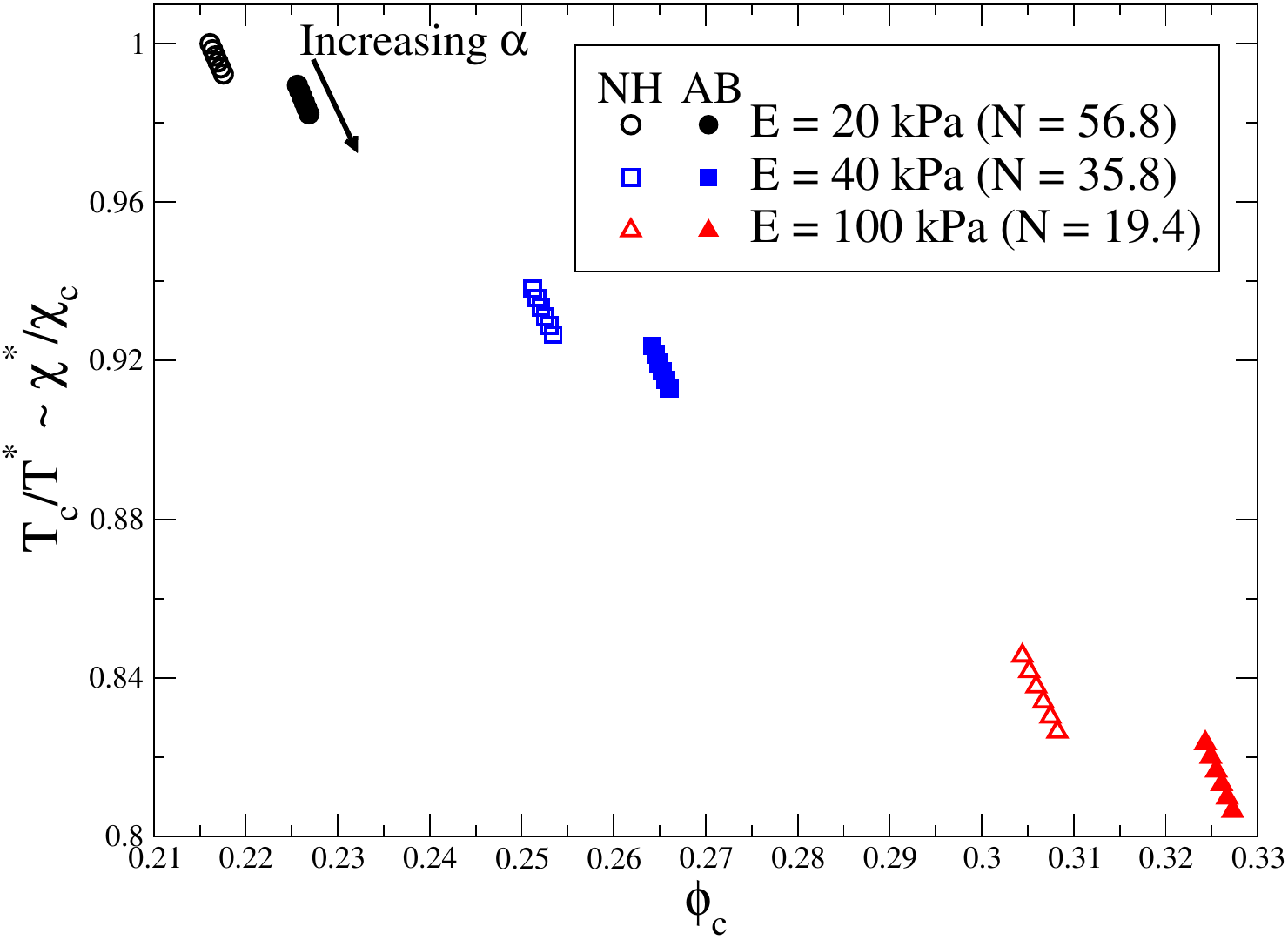}
\caption{Critical temperatures $T_c \sim 1/\chi_c$ (normalized to a temperature $T^*$ and a $\chi^*$) and critical concentration $\phi_c$ of the PDMS polymer undergoing 1D constrained-deformation deformation with the Young's moduli ranging from 20 to 100 kPa and $\alpha$ values ranging from 0 to 0.5; for the Neo-Hookean model (unfilled symbols) and the  Arruda Boyce model (filled symbols).}
\label{fig:nhab1dvara}
\end{figure}

To demonstrate the effect of $\alpha$ we use the Neo-Hookean model in 1D, for which the common tangent construction yields phase coexistence, and the AB model to study the potential importance of finite extensibility. Fig.~\ref{fig:nhab1dvara} shows the corresponding critical points for the NH (unfilled symbols) and Arruda-Boyce (filled symbols) models. For both models, the critical polymer concentration $\phi_c$ increases with increasing $\alpha$ and increasing Young's modulus $E$,  while the critical temperature $T_c\sim1/\chi_c$ decreases with them. Both phenomena are consistent with the effects of decreasing the molecular weight $N_{\textrm{eff}}$ of a polymer solution on its phase diagram \cite{doisoftmatterphys}. Although the AB model has different $\{\phi_c,T_c\}$ values than the NH model, the effect of changing $\alpha$ is similar for both models. 

We stress that a polymer gel does not have the conventional  translational `mixing' entropy $\kbT\frac1{N\text{v}_0}\phi\ln\phi$, which has the same form as the elastic osmotic term above. Despite this,  
several recent studies of LLPS used this mixing entropy in order to obtain a common tangent construction for the NH model, and calculate phase diagrams and coexistence
\cite{PhysRevX.14.021009,PhysRevLett.125.268001,D2SM01101H}. However, the attribution of this term to translational entropy of polymer strands is incorrect, and it should instead be attributed to  entropy associated with the crosslinks (if in fact $\alpha\neq0$). With our model, in a 3D deformation, one can find phase coexistence if $\alpha$ is unphysically large, even for the NH model; this would be consistent with the (unphysical) translational entropy of small polymer strands.

\section{Discussion/Conclusion}
In this paper, we applied six models to five modes of deformation to demonstrate the features of the models and the change in elastic responses with respect to deformation geometries. Two of the deformation modes, uniaxial stretch and uniaxial compression, of the incompressible materials, are the focuses of many previous studies of elasticity (especially uniaxial stretch, which is the most common method to measure a material's elastic response in experiments). The most prevalent model used in studies of elastic materials, the Neo-Hookean model does not suffice very far beyond the linear regime of the stress-strain response: if the material strain hardens, the NH model is only valid for strains $\lambda\lesssim 0.5\lambda_{\textrm{max}}$, where $\lambda_{\textrm{max}}\sim\sqrt{N}\sim (\kbT/\text{v}_0 n_{x0})^{-1/3}$ is the maximum strain of the individual polymer strands; if the material strain softens, we need an entanglement model to account for that; and if the material displays both effects, we need a combined model which would strain soften at moderate deformations and strain harden at extremely large deformations. 

By comparing our results of the incompressible material to established studies, we find that the phenomenological Mooney-Rivlin model is only applicable to a single deformation mode---uniaxial stretch. In contrast, the slip tube model, which characterizes entanglements as virtual Gaussian chains that can slip along the structural chains, reliably displays softening behavior in different modes of deformation. The two most widely used finite extensible models are the molecular Arruda-Boyce model or the simpler phenomenological FENE model. We suggest combining the Arruda-Boyce model and Rubinstein-Panyukov's slip tube model into the AB+RP model, to  incorporate both strain softening and hardening effects. 

We next studied constrained-dimension deformations in 1D, 2D, and 3D for a compressible elastic material that allows solvent transfer. In these calculations the controversial osmotic contribution (Eq.~\eqref{eq:fingel} or \eqref{eq:fosm_N}) to the elastic free energy is important, but is normally overlooked. In compressible extension or swelling, increases in elastic energy are countered by decreases in free energy density as a result of the volume increase, resulting in different elastic responses in the volume-changing deformations than in incompressible deformations, depending on the dimension D of the deformation. The most substantial effect of this competition is that the common tangent construction fails when applied to the commonly-used neo-Hookean model, undergoing a 3D deformation.

Because most gel phase separation occurs in 3D, one proposal has been to use a more complex concentration-dependent interaction parameter $\chi(\phi)$ \cite{TF9716702275, 10.1063/1.460672, ermanFlory1986} to obtain the `W' shaped free energy curve to that permits a common tangent construction. However, others have circumvented the problem by using the Gent model for cavity growth and assuming the cavity has a sharp interface \cite{doi:10.1098/rspa.1959.0016,Horgan:2004aa, HORGAN1992279, Ronceray_2022, KOTHARI2020104153, PhysRevE.107.024418, RAAYAIARDAKANI2019100536}. In this paper, we propose using finite extensible models such as the Arruda-Boyce model and the FENE model to obtain a continuous network during phase separation. We showed with the phase diagrams that the deformation dimensions would significantly affect the critical behavior of an elastic material. In addition, we show how the controversial volume-sensitive term $\alpha \ln I_3$ affects phase behavior. 
\section*{Author contributions}
\textbf{Conceptualization, Methodology, Interpretation, Writing (Original Draft), Writing (review \& editing):} SW and PDO;
\textbf{Software, Formal Analysis, Calculations, Analysis:} SW;
\textbf{Resources, Supervision:} PDO. 

\section*{Conflicts of interest}
There are no conflicts to declare.

\section*{Data availability}

Data plotted in all figures can be calculated directly from the formulas provided. 

\section*{Acknowledgments}

We thank Eric Dufresne, Carla Fernandez Rico, and Rob Style for helpful discussions. PDO is grateful to Georgetown University and the Ives Foundation for support.


\appendix
\setcounter{figure}{0}
\renewcommand{\thefigure}{A\arabic{figure}}

\section{Maximum stretch of the finite extensible models} \label{appx:maxl}
For a network that deforms from the relaxed state according to Gaussian statistics where $\langle R^2\rangle_{0i} = Na^2/3$, its deformed state satisfies $\langle R^2\rangle = I_1 \cdot (Na^2/3)$. The maximum end-to-end distance of $N$ Kuhn steps polymer chain is $R_{\text{max}} = Na$ or $R_{\text{max}} = (Na)^2$, so the invariant $I_1$ must satisfy $I_1 \leq 3N$ for finite extensible models. The maximum extension ratios for incompressible deformations are therefore:
\begin{align}
\lambda_{\text{max/min}}^2 + \frac{2}{\lambda_{\text{max/min}}} &= 3N \indent &&\textrm{(uniaxial deformation)}\\
\lambda_0^2(\text{D}\lambda_{\text{max}}^2 + 3-\text{D}) &= 3N \indent &&\textrm{(D-dimension constrained)}
\end{align}
where for the uniaxial deformation, $\lambda_\text{max}>1$ is the maximum stretch ratio in the loading direction for a uniaxial stretch, and $\lambda_\text{min}<1$ is the minimum (compressive) stretch ratio in the loading direction for an uniaxial compression; And for the constrained dimension deformation, $\lambda_0$ is the isotropic pre-stretch, and $\lambda_\text{max}$ is in the unconstrained dimension(s). The theoretical minimum stretch is 0 if we ignore excluded volume and other intermonomer interactions.

\section{Concentration dependent $\chi(T,\phi)$} \label{appx:varchi}
\begin{figure*}
\includegraphics[width=0.9\textwidth]{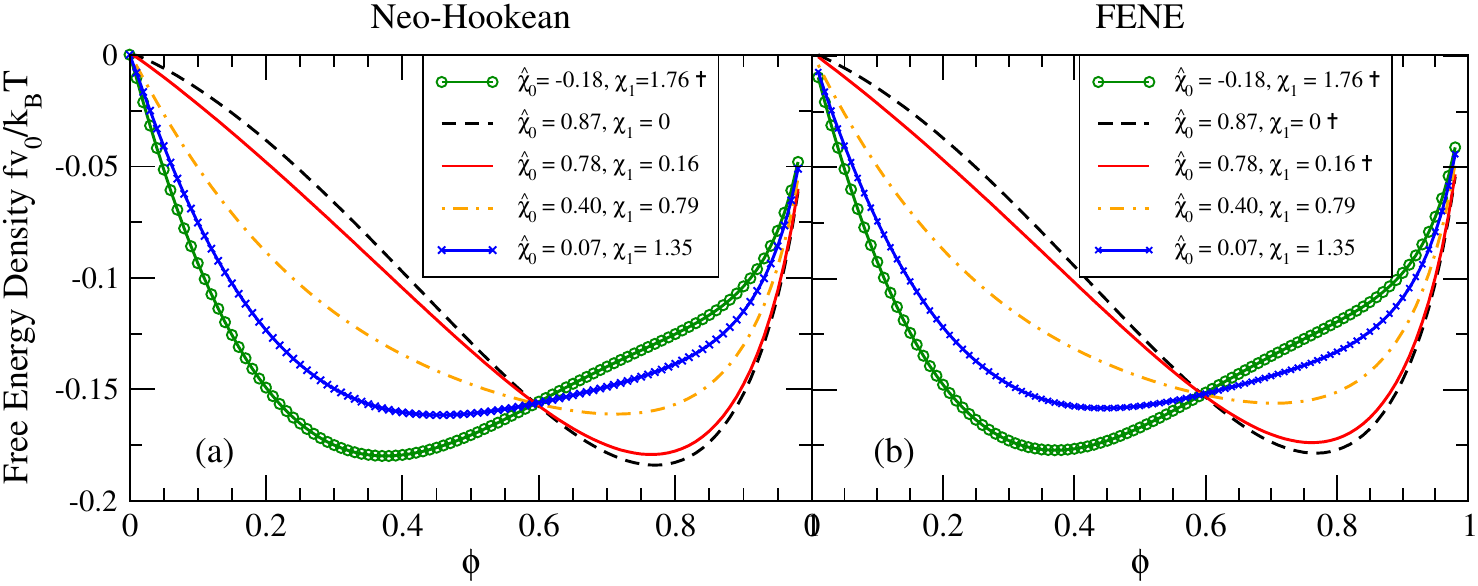}
\caption{(a) Neo-Hookean and (b) FENE model of 3D deformation with a concentration dependent mixing parameter $\chi$. The parameters used here are $E = 40$ kPa, $\phi_\text{sw} = 0.595$, $\alpha_\text{NH} = 0.5$, and $\alpha_\text{FENE} = 0.529$. Values that permit the common tangent construction are labeled with \textdagger \;sign. For the NH model (left) it is only possible with the green circled line, whose $\chi_1/\hat{\chi}_0$ value is not common with a PDMS gel. The FENE model (right) has a common tangent for more realistic values of $0\leq\chi_1/\hat{\chi}_0<1$, thanks to its finite extensibility.}
\label{fig:varchi}
\end{figure*}

The concentration dependence of the Flory parameter $\chi(T,\phi)$ can be empirically expanded in powers of concentration $\phi$ as \cite{10.1063/1.460672,TF9716702275,ermanFlory1986}
\begin{equation} \label{eq:chiexpand}
\chi = \hat{\chi}_0(T) +\chi_1 \phi + \chi_2 \phi^2+ \dots,
\end{equation}
where $\left\{\chi_i\right\}$ can be determined from osmotic and vapor pressure measurements \cite{doi:10.1021/ja01517a022} of the uncrosslinked polymer solution. In this section, we will keep up to the linear term in $\phi$ so that $\chi_i = 0$ for $i>1$. In principle, the higher order terms arise from excluded volume and other specific interactions \cite{ScalingDeGenne}. Since $\chi$ depends on the species of the mixture and the data might not be available, it is beneficial to study what range of $\hat{\chi}_0$ and $\chi_1$ can enable a common tangent construction. Fig.~\ref{fig:varchi} shows the free energy densities calculated for the NH and FENE models within a 3D deformation (for which the NH model does not provide a common tangent construction for $\chi_1=0$, as shown in Fig.~\ref{fig:pd3dim} in the main text). The values of $\hat{\chi}_0, \chi_1$ are chosen to ensure that the saturation concentration $\phi_{\text{sw}}$ is the same and the ratio of $\chi_1/\hat{\chi}_0$ varies from -10 to 20. The NH model cannot support a common tangent construction for a positive small ratio $\chi_1/\hat{\chi}_0$, consistent with the discussion in Section.~\ref{subsec:fe&str}. The lack of a common tangent construction was discussed by \citet{10.1063/1.460672}, who proposed using the concentration dependent $\chi(\phi)$ as a resolution. We find here that only specific values of $\hat{\chi}_0$ and $\chi_1$ can support a phase separation with non-finite extensible elastic model (in Fig.~\ref{fig:varchi}, only the green circled line supports a common tangent construction). In other words, whether the network-solvent mixture can undergo a phase separation depends on the chemistry of the component, unlike our approach with the finite extensible elastic models, which guarantees a phase separation.

For the FENE model, when the ratio of $\chi_1/\hat{\chi}_0$ is relatively small, \textit{i.e.} $0 \leq \chi_1/\hat{\chi}_0 <1$, the free energy density does not differ much from each other. \citet{ermanFlory1986} showed that for PS in cyclohexane and PIB in benzene, the ratios are in such a range $0 \leq \chi_1/\hat{\chi}_0 <1$. More importantly, experiments with PDMS and HFBMA solution \cite{Carla_unpublished_work} showed that for the materials that concern our study, $\chi_1/\hat{\chi}_0 \simeq 0.3$, which is in this range of small $\chi_1/\hat{\chi}_0$. As a result, using a more realistic concentration dependent $\chi(T,\phi)$ function does not result in a very different free energy than just using the simplest case where $\chi_1 = 0$ (plotted in black). We therefore chose to ignore the concentration dependency of the $\chi$ parameter in this paper to focus on the effect of elasticity. Note that we also ignore the probability that $\hat{\chi}_0(T)$ and $\chi_1(T)$ themselves are functions of temperature in order to make a naive characterization of the phase transition with a concentration dependent Flory parameter $\chi(T,\phi)$. 

\balance


\bibliography{Reference} 
\bibliographystyle{rsc} 
\end{document}



\title{Supplementary Information for ``The roles of elasticity and dimension in liquid-gel phase separation''}



\author{Shichen Wang and Peter D. Olmsted} 


\maketitle

\section{Free energy densities}
In this section, we summarize the explicit expressions for the free energy densities and stresses of different models with various modes of deformation.
\subsection{Incompressible uniaxial deformations}
\noindent Neo-Hookean model:
\begin{align} 
f^0_{\text{NH}} &= \frac{\Phi}{4}Cn^\text{NH}_{x0}k_B T  \left(\lambda^2+\frac{2}{\lambda} -3 \right), \label{eq:n-h-b}\\
\sigma_{zz} &= \sigma_{zz}^{\text{NH}} - \sigma_{xx}^{\text{NH}}= \frac{\Phi}{2}Cn^\text{NH}_{x0}k_B T  \left(\lambda^2-\frac{1}{\lambda}\right).  \label{eq:n-h-s-b}
\end{align}
Arruda-Boyce model:
\begin{align}
f^0_{\text{AB}} &= \frac{\Phi}{2}Cn^\text{AB}_{x0}k_B T\left[ \sqrt{N}\beta\sqrt{\frac{\lambda^2 + 2/\lambda}{3}} - 
 N \ln \left(\frac{\sinh \beta}{\beta}\right)\right],  \label{eq:a-b-b}\\
\sigma_{zz} &= \sigma_{zz}^{\text{AB}} - \sigma_{xx}^{\text{AB}}= \frac{\Phi}{2}Cn^\text{AB}_{x0}k_B T\sqrt{\frac{N}{3}}\beta\frac{\lambda^2-\lambda^{-1}}{\sqrt{\lambda^2 + 2/\lambda}} ,\label{eq:ab-s-b}
\end{align}
where
\begin{equation} \label{eq:langevin-b}
\coth \beta - \frac{1}{\beta} =\sqrt{\frac{\lambda^2 + \frac{2}{\lambda}}{3N}}.
\end{equation}
FENE model:
\begin{align} 
f^0_{\text{FENE}} &= \frac{\Phi}{4} C n^\text{FN}_{x0} k_B T\left[-3N\text{ln}\left(1-\frac{\lambda^2+\frac{2}{\lambda}}{3N}\right) \right], \label{eq:fene-b}\\
\sigma_{zz} &= \sigma_{zz}^{\text{FENE}} - \sigma_{xx}^{\text{FENE}} = \frac{\Phi}{2}C n^\text{FN}_{x0} k_B T\frac{\lambda^2 - \lambda^{-1}}{1-\left(\lambda^2+\frac{2}{\lambda}\right)/3N}.\label{eq:fene-s-b}
\end{align}
Mooney-Rivlin model:
\begin{align} 
f^0_{\text{MR}} &=  C_1(\lambda^2+\frac{2}{\lambda}-3)+C_2(\frac{1}{\lambda^2}+2\lambda-3), \label{eq:m-r-b}\\ 
t_{zz} &= \sigma_{zz}^{\text{MR}} - \sigma_{xx}^{\text{MR}}=  2C_1(\lambda^2-\frac{1}{\lambda})+2C_2(\lambda-\frac{1}{\lambda^2}).\label{eq:mr-s-b}
\end{align}
Rubinstein Panyukov and ABRP model:
\begin{align} 
\frac{f^0_{\text{RP}}}{\frac{\Phi}{4}n^\text{RP}_{x0}k_BT} &= C\left(\lambda^2+\frac{2}{\lambda}\right)+L\left[\frac{\lambda}{\sqrt{g}} + \frac{\sqrt{g}}{\lambda} +\frac{2}{\sqrt{\lambda}}\sqrt{\frac{2}{3-g}} + \sqrt{2\lambda(3-g)} \right]-\frac{2L}{3} \ln \left[\left(\frac{N}{L}\right)^3 g\left(\frac{3-g}{2}\right)^2 \right], \label{eq:s-t-b} \\
t_{zz} &= \sigma_{zz}^{\text{RP}} - \sigma_{xx}^{\text{RP}}=  \frac{\Phi}{4} n^\text{RP}_{x0} k_B T \Bigg\{2C\left(\lambda^2-\frac{1}{\lambda}\right)+L\left[\frac{\lambda}{\sqrt{g}} - \frac{\sqrt{g}}{\lambda} -\sqrt{\frac{2}{\lambda (3-g)}} + \sqrt{\frac{\lambda(3-g)}{2}} \right]\Bigg\}, \label{eq:rp-s-b} \\
\begin{split} 
\frac{f^0_{\text{ABRP}}}{\frac{\Phi}{4}n^\text{ABRP}_{x0} k_B T } &=  2C\left[ \sqrt{N}\beta\sqrt{\frac{\lambda^2 + 2/\lambda}{3}} - N \ln \left(\frac{\sinh \beta}{\beta}\right)\right] + L\left[\frac{\lambda}{\sqrt{g}} + \frac{\sqrt{g}}{\lambda} +\frac{2}{\sqrt{\lambda}}\sqrt{\frac{2}{3-g}} + \sqrt{2\lambda(3-g)} \right] \\
& \;\;\;\;\;\;-\frac{2L}{3} \ln \left[\left(\frac{N}{L}\right)^3 g\left(\frac{3-g}{2}\right)^2 \right],
\end{split} \label{eq:abst-in} \\
\begin{split} 
t_{zz} &= \sigma_{zz}^{\text{ABRP}} - \sigma_{xx}^{\text{ABRP}} \\
&=  \frac{\Phi n^\text{ABRP}_{x0} k_B T}{4} \Bigg\{2C\sqrt{\frac{N}{3}}\beta\frac{\lambda^2-\lambda^{-1}}{\sqrt{\lambda^2 + 2/\lambda}}+L\left[\frac{\lambda}{\sqrt{g}} - \frac{\sqrt{g}}{\lambda} -\sqrt{\frac{2}{\lambda (3-g)}} + \sqrt{\frac{\lambda(3-g)}{2}} \right]\Bigg\},  \end{split} \label{eq:abrp-s-b}
\end{align}
where the parameter $g$ is given by:
\begin{equation} \label{eq:g-uni}
-\frac{\lambda}{g^{3/2}}+\frac{1}{\lambda\sqrt{g}} - \frac{4}{3g} = -\frac{1}{\sqrt{\lambda}}\left(\frac{2}{3-g}\right)^{3/2} + \sqrt{\frac{2\lambda}{3-g}} - \frac{8}{3(3-g)}, 
\end{equation}
and $\beta$ is given by Eq.~\eqref{eq:langevin-b}.

\begin{figure*}
\includegraphics[width=0.9\textwidth]{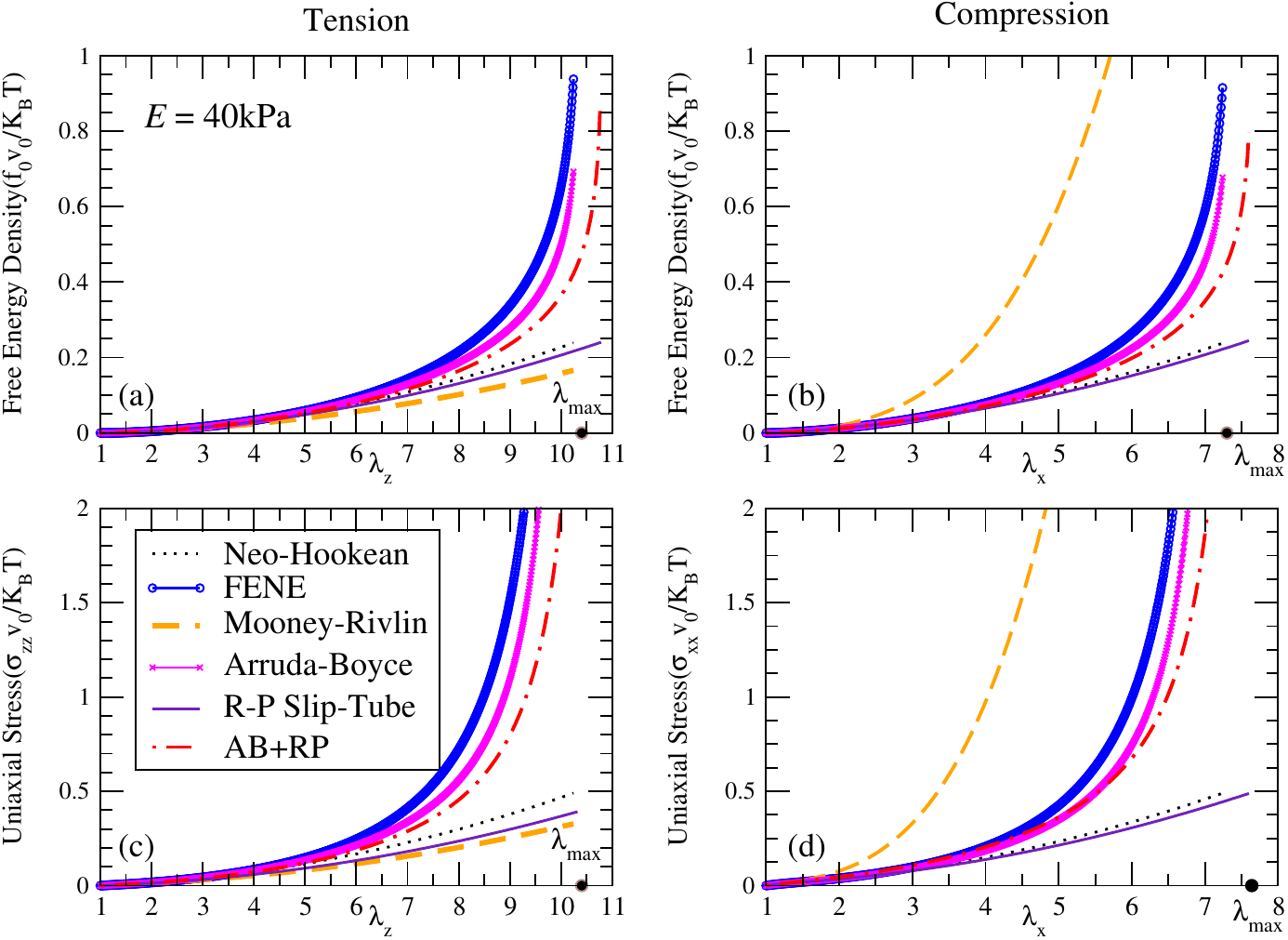}
\caption{Elastic free energy and stress of an incompressible PDMS network undergoing uniaxial tension (a, c) and compression (b, d). }
\label{fig:modelsES}
\end{figure*}

We show the free energy densities and the true stresses of the six models in Fig.~\ref{fig:modelsES} for uniaxial compression and extension, in addition to the Mooney ratio plot shown in Fig.~3 in the main text.

We show in the main text that the Mooney ratio $t^M = t_{zz}/(\lambda^2-\lambda^{-1})$ for the slip tube model does not monotonically decrease in uniaxial compression; instead, as $\lambda_{zz}$ decreases below unity (so that $\lambda_{xx} = \lambda_{yy} = \lambda^{-1/2}_{zz}$ in the transverse directions increases), the Mooney ratio increases first, signaling strain hardening, and then decreases, signaling strain softening. Mathematically, this is an intrinsic feature of the slip tube model, because $t^M \sim 2C + L (\lambda/\sqrt{g} - \sqrt{g}/\lambda - \cdots )/(\lambda^2 - \lambda^{-1})$ has its maximum at 
\begin{equation} \label{eq:dLdlambda}
\frac{d}{d\lambda}\left\{\frac{1}{\lambda^2-\lambda^{-1}}\left[\frac{\lambda}{\sqrt{g}} - \frac{\sqrt{g}}{\lambda} -\sqrt{\frac{2}{\lambda (3-g)}} + \sqrt{\frac{\lambda(3-g)}{2}} \right]\right\} = 0,
\end{equation}
whose solution is $\lambda= 0.558$, independent of any parameter of the network. Note that $g=g(\lambda)$ depends on the deformation according to Eq.~\ref{eq:g-uni}, and, in general, a function $\bm{f}(\lambda, g)$ would have its derivative with respect to $\lambda$ in the form
\begin{equation} \label{eq:dfdlambda}
    \frac{d\bm{f}(\lambda,g)}{d\lambda} = \frac{\partial \bm{f}}{\partial \lambda} + \frac{\partial g}{\partial \lambda}\frac{\partial \bm{f}}{\partial g},
\end{equation}
where $\frac{\partial g}{\partial \lambda}$ can be found by taking partial derivatives of Eq.~\ref{eq:g-uni}. However, when the function is the elastic free energy density, $\bm{f}(\lambda,g) = f^0_\text{RP}$, we can use the condition that the free energy is minimized with respect to $g$, \textit{i.e.} $\frac{\partial f^0_\text{RP}}{\partial g} = 0$, to simplify the calculation by only taking the partial derivative with respect to $\lambda$. This is not the case for other functions, and the derivatives have to include both partial derivatives in Eq.~\eqref{eq:dfdlambda}.

\subsection{Volume-changing deformations (elasticity)} \label{appx:vcd}
In general, when we calculate the stress tensor $\sigma_{ii}$ from the free energy density $f$, we need to treat the deformation $\lambda_i$ in each direction as an independent variable, so that 
\begin{align}
\sigma_{ii} &= f\delta_{ii}+\lambda_i \left.\frac{\partial f}{\partial \lambda_{i}}\right|_{\lambda_{j\neq i}},\label{eq:tstresscomp_diagA} \\
& = f\delta_{ii}+\lambda_i \left.\left(\frac{\partial f}{\partial I_1} \frac{\partial I_1}{\partial \lambda_i} +\frac{\partial f}{\partial I_2} \frac{\partial I_2}{\partial \lambda_i} +\frac{\partial f}{\partial I_3} \frac{\partial I_3}{\partial \lambda_i}  \right)\right|_{\lambda_{j\neq i}}, \label{eq:tstresscomp_diagB}\\
& = f\delta_{ii}+\lambda_i \left.\left[\frac{\partial f}{\partial I_1} 2\lambda_i +\frac{\partial f}{\partial I_2} 2\lambda_i(I_1 - \lambda_i^2) +\frac{\partial f}{\partial I_3} 2 \frac{I_3}{\lambda_i}  \right]\right|_{\lambda_{j\neq i}}; \label{eq:tstresscomp_diagC}
\end{align}
or from the free energy density in the reference state $f^0$:
\begin{align}
\sigma_{ii} &= f\delta_{ii}+\lambda_i \left.\frac{\partial \frac{f^0}{\sqrt{I_3}}}{\partial \lambda_{i}}\right|_{\lambda_{j\neq i}},\label{eq:tstresscomp_diagD} \\
& = f\delta_{ii} - \lambda_i \frac{1}{2} I_3^{-3/2} f^0 \left.\frac{\partial I_3}{\partial \lambda_{i}}\right|_{\lambda_{j\neq i}} + \frac{1}{\sqrt{I_3}}\left.\frac{\partial f^0}{\partial \lambda_i}\right|_{\lambda_{j\neq i}}, \label{eq:tstresscomp_diagE}\\
& = f\delta_{ii} - I_3^{-1/2} f^0 \delta_{ii}+\frac{1}{\sqrt{I_3}}\left.\frac{\partial f^0}{\partial \lambda_i}\right|_{\lambda_{j\neq i}}, \label{eq:tstresscomp_diagF}\\
& = \frac{1}{\sqrt{I_3}}\left.\frac{\partial f^0}{\partial \lambda_i}\right|_{\lambda_{j\neq i}}, \label{eq:tstresscomp_diagG}\\
& = \frac{2}{\sqrt{I_3}} \left.\left[\frac{\partial f^0}{\partial I_1} \lambda^2_i +\frac{\partial f^0}{\partial I_2} \lambda^2_i(I_1 - \lambda_i^2) +\frac{\partial f}{\partial I_3} I_3 \delta_{ii}\right]\right|_{\lambda_{j\neq i}}.\label{eq:tstresscomp_diagH}
\end{align}

From the above derivation we can present the elastic energy cost for performing a constrained-dimension deformation of the forms 
\begin{align}\label{eq:lambdadB}
\vec{\lambda}=\begin{cases}
(\lambda,1,1)& (\text{D}=1) \\
(\lambda,\lambda,1)& (\text{D}=2) \\
(\lambda,\lambda,\lambda)& (\text{D}=3).
\end{cases}
\end{align}

The stress in the constrained direction $\lambda_i = 1$ is 
\begin{equation} \label{eq:tstresscomp_usA}
\sigma_{ii} = \frac{2}{\sqrt{I_3}}\left[\frac{\partial f^0}{\partial I_1} +\frac{\partial f^0}{\partial I_2}(I_1 - 1) +\frac{\partial f}{\partial I_3} I_3 \right] \delta_{ii} = \frac{2}{\sqrt{I_3}}(f^0)'.
\end{equation}
In the simplest case where the free energy only depends on composition $f(\phi)$, we can use $f^0=f\phi_\text{id}/\phi= f/\phi$ and $\sqrt{I_3}=\phi_\text{id}/\phi = \phi^{-1}$ to show that $\sigma_{ii}=f-\phi\partial f/\partial\phi = -\Pi_\text{osm}$.

We then directly use $\lambda_i = \lambda$ in the constrained direction to find their principal stresses. In this case, because we can have the same extension ratio in multiple directions ($\lambda_i =\lambda_j =\dots= \lambda$), the stress(es) in the unconstrained direction(s) depend(s) on the deformation dimension D according to Eq.~\eqref{eq:tstresscomp_diagC} or~\eqref{eq:tstresscomp_diagH}, because the direct derivatives would result in extra pre-factors of 2 or 3 by double/triple counting the stresses. For example, for D $=2$, 
 \begin{equation}
\left.\lambda_x\frac{\partial f(\lambda_x,\lambda_y,\lambda_z)}{\partial\lambda_x}\right|_{\lambda_x=\lambda_y=\lambda,\lambda_z=1}=\frac12\frac{\partial f}{\partial\lambda},
\end{equation}
where the derivative $\partial/\partial\lambda$ is to be taken after replacing $\vec{\lambda}=(\lambda,\lambda,1)$. 
This allows us to calculate the stress components $\sigma_{jj}$ in the directions where the deformation occurs. Hence we find
\begin{subequations}
\begin{align}
\sigma_{xx} &= f+\lambda \frac{\partial f}{\partial \lambda},\quad \sigma_{yy}=\sigma_{zz}=\frac{2}{\sqrt{I_3}}(f^0)'&&(1\text{D})\\
\sigma_{xx} &= \sigma_{yy} =f+\frac{\lambda}{2} \frac{\partial f}{\partial \lambda}, \quad\sigma_{zz}=\frac{2}{\sqrt{I_3}}(f^0)'&&(2\text{D}) \\
\sigma_{xx} &= \sigma_{yy}=\sigma_{zz} = f+\frac{\lambda}{3} \frac{\partial f}{\partial \lambda}. &&(3\text{D})
\end{align}
\end{subequations}

We can write the free energies and stresses in the unconstrained direction for the six models in terms of the deformation dimension D:
\begin{align} 
\frac{f^0_{\text{NH}}}{(\Phi/4)n^\text{NH}_{x0} k_B T} &= C\left(\text{D}\lambda^2-\text{D}\right) - 2\text{D}\alpha\ln\lambda, \label{eq:nh-c} \\
\frac{\sigma_{\text{NH}}}{(\Phi/4)n^\text{NH}_{x0}k_B T} &= 2C\lambda^2 - 2\alpha,  \label{eq:nh-s-c} \\
\frac{f^0_{\text{AB}}}{(\Phi/4)n^\text{AB}_{x0}k_B T}&=2C\left[\sqrt{N}\beta\sqrt{\frac{\text{D}\lambda^2 + 3-\text{D}}{3}} - 
 N \ln \left(\frac{\sinh \beta}{\beta}\right)\right]- 2\text{D}\alpha\ln\lambda,  \label{eq:a-b-c} \\
\frac{\sigma_{\text{AB}}}{(\Phi/4)n^\text{AB}_{x0}k_B T} &= 2C \sqrt{\frac{N}{3}}\beta \frac{\lambda^2}{\sqrt{\text{D}\lambda^2 + 3-\text{D}}}- 2\alpha, \label{eq:ab-s-c} \\
\frac{f^0_{\text{FENE}}}{(\Phi/4)n^\text{FN}_{x0} k_B T}& = C\left[-3N\text{ln}\left(1-\frac{\text{D}\lambda^2 + 3-\text{D}}{3N}\right)\right]- 2\text{D}\alpha\ln\lambda, \label{eq:fene-c}\\
\frac{\sigma_{\text{FENE}}}{(\Phi/4)n^\text{FN}_{x0} k_B T} &= 2C\frac{\lambda^2}{1-\left(\text{D}\lambda^2+3-\text{D}\right)/3N}- 2\alpha,  \label{eq:fene-s-c}
\end{align}
 where
\begin{equation} \label{eq:langevin-c}
\coth \beta - \frac{1}{\beta} =\sqrt{\frac{\text{D}\lambda^2 + 3-\text{D}}{3N}}
\end{equation}
\begin{align} 
f^0_{\text{MR}} &= 
\begin{cases}
C_1(\lambda^2-1)+C_2(2\lambda^2-2)-\frac{\Phi}{2}n^\text{MR}_{x0}k_BT\alpha \ln \lambda& \textrm{(1D)} \\[10truept] 
C_1(2\lambda^2-2)+C_2( \lambda^4 + 2 \lambda^2-3) -\Phi n^\text{MR}_{x0}k_BT \alpha \ln \lambda& \textrm{(2D)}\\[10truept] 
C_1(3\lambda^2-3)+C_2(3 \lambda^4 -3)- \frac{3\Phi}{2} n^\text{MR}_{x0}k_BT\alpha \ln \lambda& \textrm{(3D)}
\end{cases} \label{eq:mr-c} \\[10truept]
\sigma_{\text{MR}} &= 
\begin{cases}
2C_1\lambda^2+4C_2\lambda^2- \frac{\Phi}{2}n^\text{MR}_{x0}k_BT\alpha& \textrm{(1D)} \\[10truept] 
2C_1\lambda^2+ 2C_2( \lambda^4 + \lambda^2) - \frac{\Phi}{2} n^\text{MR}_{x0}k_BT\alpha& \textrm{(2D)}\\[10truept] 
2C_1\lambda^2+4 C_2\lambda^4 - \frac{\Phi}{2} n^\text{MR}_{x0}k_BT\alpha& \textrm{(3D)}
\end{cases} \label{eq:mr-s-c}
\end{align}
\begin{align} 
&\frac{f^0_{\text{RP}}}{(\Phi/4)n^\text{RP}_{x0} k_B T}= &&C\left(\text{D}\lambda^2+3-\text{D}\right)+L\left[\text{D}\left(\frac{\lambda}{\sqrt{g}} + \frac{\sqrt{g}}{\lambda}\right) +(3-\text{D})\left(\sqrt{\frac{3-\text{D}}{3-\text{D}g}} + \sqrt{\frac{3-\text{D}g}{3-\text{D}}}\right) \right]\nonumber\\
&&&-\frac{2L}{3} \ln \left[\left(\frac{N}{L}\right)^3 g^\text{D}\left(\frac{3-\text{D}g}{3-\text{D}}\right)^{3-\text{D}} \right]- 2\text{D}\alpha\ln\lambda,\label{eq:s-t-c} \\
&\frac{\sigma_{\text{RP}}}{(\Phi/4)n^\text{RP}_{x0} k_B T}= &&2C\lambda^2+L\left(\frac{\lambda}{\sqrt{g}} - \frac{\sqrt{g}}{\lambda}\right) - 2\alpha, \label{eq:st-s-c} \\
&\frac{f^0_{\text{ABRP}}}{(\Phi/4)n^\text{ABRP}_{x0} k_B T} = &&L\left[\text{D}\left(\frac{\lambda}{\sqrt{g}} + \frac{\sqrt{g}}{\lambda}\right) +(3-\text{D})\left(\sqrt{\frac{3-\text{D}}{3-\text{D}g}} + \sqrt{\frac{3-\text{D}g}{3-\text{D}}}\right) \right]- 2\text{D}\alpha\ln\lambda \label{eq:abrp-c}\\
&&& +2CN\left[\beta\sqrt{\frac{\text{D}\lambda^2 + 3-\text{D}}{3N}} -\ln \left(\frac{\sinh \beta}{\beta}\right)\right] -\frac{2L}{3} \ln \left[\left(\frac{N}{L}\right)^3 g^\text{D}\left(\frac{3-\text{D}g}{3-\text{D}}\right)^{3-\text{D}} \right], \nonumber \\
&\frac{\sigma_{\text{ABRP}}}{(\Phi/4)n^\text{ABRP}_{x0} k_B T}= &&2C\sqrt{\frac{N}{3}}\beta\frac{\lambda^2}{\sqrt{\text{D}\lambda^2+3-\text{D}}}+L\left(\frac{\lambda}{\sqrt{g}} - \frac{\sqrt{g}}{\lambda}\right) - 2\alpha, \label{eq:abrp-s-c} 
\end{align}
and $\beta$ is given by Eq.~\eqref{eq:langevin-c}.

\subsection{Gel Free Energies}
Here we show free energy densities and stresses in the unconstrained direction(s) as measured in the current (deformed) frame, written in terms of the polymer volume fraction $\phi$.
\vskip0.5truecm
\noindent Neo-Hookean model:
\begin{align} 
\frac{f_{\text{NH}}}{ (\Phi/4)n^\text{NH}_{x0} k_B T } &=\phi \left\{C\phi_{\text{sw}}^{-2/3}\left[\text{D}\left(\frac{\phi_{\text{sw}}}{\phi}\right)^{2/\text{D}}+3-\text{D}\right]-3C +2\alpha \ln \phi\right\} \label{eq:n-h-d}\\
\frac{\sigma_{\text{NH}}}{(\Phi/4)n^\text{NH}_{x0} k_B T} &= \phi\left[2C \phi_{\text{sw}}^{\frac{2}{\text{D}}-\frac{2}{3}}\phi^{-\frac{2}{\text{D}}}-2\alpha \right] \label{eq:nh-s-d}
\end{align}
Arruda-Boyce model:
\begin{align} 
\frac{f_{\text{AB}}}{(\Phi/4)n^\text{AB}_{x0} k_B T}&=\phi\left\{2C\left[\sqrt{N}\beta\sqrt{\frac{\text{D}(\phi_{\text{sw}}/\phi)^{2/\text{D}} + 3-\text{D}}{3\phi_{\text{sw}}^{2/3}}} - 
 N \ln \left(\frac{\sinh \beta}{\beta}\right)\right] +2\alpha \ln\phi\right\} \label{eq:a-b-d}\\
\frac{\sigma_{\text{AB}}}{(\Phi/4)n^\text{AB}_{x0} k_B T}&=\phi\left[2C\sqrt{\frac{N}{3\phi_{\text{sw}}^{2/3}}}\left(\frac{\phi_{\text{sw}}}{\phi}\right)^{2/\text{D}}\frac{\beta}{\sqrt{\text{D}(\phi_{\text{sw}}/\phi)^{2/\text{D}} + 3-\text{D}}} -2\alpha \right] \label{eq:ab-s-d}
\end{align}
where
\begin{equation} \label{eq:langevin-d}
\coth \beta - \frac{1}{\beta} =\sqrt{\frac{\text{D}(\phi_{\text{sw}}/\phi)^{2/\text{D}} + 3-\text{D}}{3\phi_{\text{sw}}^{2/3}N}}
\end{equation}
FENE model:
\begin{align}
\frac{f_{\text{FENE}}}{(\Phi/4)n^\text{FN}_{x0} k_B T} &=  \phi\left\{C\left[-3N\text{ln}\left(1-\frac{\text{D}\left(\frac{\phi_{\text{sw}}}{\phi}\right)^{\frac{2}{\text{D}}} + 3-\text{D}}{3\phi_{\text{sw}}^{2/3}N}\right)\right] +2\alpha \ln \phi\right\}  \label{eq:fene-d}\\
\frac{\sigma_{\text{FENE}}}{(\Phi/4)n^\text{FN}_{x0} k_B T}&=  \phi \left[\frac{2C\left(\frac{\phi_{\text{sw}}}{\phi}\right)^{2/\text{D}}}{\phi_{\text{sw}}^{2/3}-\left(\text{D}\left(\frac{\phi_{\text{sw}}}{\phi}\right)^{\frac{2}{\text{D}}} + 3-\text{D}\right)/3N} -2\alpha \right]\label{eq:fene-s-d}
\end{align}
Mooney-Rivlin model:
\begin{align} 
    f_{\text{MR}}&= 
    \begin{cases}
    \phi\left\{C_1\left[\frac{1}{\phi_{\text{sw}}^{2/3}}\left(\frac{\phi_{\text{sw}}^2}{\phi^2}+2\right) -3\right] + C_2\left[\frac{1}{\phi_{\text{sw}}^{4/3}}\left(2\frac{\phi_{\text{sw}}^2}{\phi^2}+1\right) -3\right]+ \frac{\Phi}{2}n^\text{MR}_{x0}k_B T \alpha \ln \phi\right\}& \textrm{(1D)} \\[10truept] 
    \phi\left\{C_1\left[\frac{1}{\phi_{\text{sw}}^{2/3}}\left(2\frac{\phi_{\text{sw}}}{\phi}+1\right) -3\right]+C_2\left[ \frac{1}{\phi_{\text{sw}}^{4/3}}\left(\frac{\phi_{\text{sw}}^2}{\phi^2} + 2 \frac{\phi_{\text{sw}}}{\phi}\right)-3\right] +\frac{\Phi}{2}n^\text{MR}_{x0}k_B T \alpha \ln \phi\right\}& \textrm{(2D)}\\[10truept] 
    \phi\left[3C_1\left(\frac{1}{\phi^{\frac{2}{3}}}-1\right)+3C_2\left(\frac{1}{\phi^{\frac{4}{3}}} -1\right)+\frac{\Phi}{2}n^\text{MR}_{x0}k_B T \alpha \ln \phi\right]& \textrm{(3D)}
    \end{cases} \label{eq:m-r-d}\\
    \nonumber \\
   \sigma_{\text{MR}} &= 
    \begin{cases}
    2\left[\left(C_1+ 2C_2\right)\phi_{\text{sw}}^{\frac{4}{3}}\phi^{-2}- \frac{\Phi}{4}n^\text{MR}_{x0}k_B T\alpha \right]\phi & \textrm{(1D)} \\[10truept] 
    2\left[C_1\phi_{\text{sw}}^{\frac{1}{3}}\phi^{-1}+C_2 \left(\phi_{\text{sw}}^{\frac{2}{3}}\phi^{-2} +\phi_{\text{sw}}^{-\frac{1}{3}}\phi^{-1}\right) - \frac{\Phi}{4}n^\text{MR}_{x0}k_B T\alpha\right]\phi& \textrm{(2D)}\\[10truept] 
    2\left[C_1\phi^{-\frac{2}{3}}+2C_2\phi^{-\frac{4}{3}} -\frac{\Phi}{4}n^\text{MR}_{x0}k_BT\alpha \right]\phi& \textrm{(3D)}
    \end{cases}\label{eq:mr-s-d}
\end{align}
Rubinstein-Panyukov:
\begin{align} 
&\frac{f_{\text{RP}}}{(\Phi/4)n^\text{RP}_{x0} k_B T} &&= \phi \Bigg\{L\left[\text{D}\left(\frac{\phi_{\text{sw}}^{\frac{1}{\text{D}}-\frac{1}{3}}}{\sqrt{g}\phi^{\frac{1}{\text{D}}}} +\frac{\sqrt{g}\phi^{\frac{1}{\text{D}}}}{\phi_{\text{sw}}^{\frac{1}{\text{D}}-\frac{1}{3}}} \right)+(3-\text{D})\left(\frac{1}{\phi_{\text{sw}}^{\frac{1}{3}}}\sqrt{\frac{3-\text{D}}{3-\text{D}g}} + \phi_{\text{sw}}^{\frac{1}{3}}\sqrt{\frac{3-\text{D}g}{3-\text{D}}} \right)\right]\nonumber \\
&&& \indent \indent + C\phi_{\text{sw}}^{-2/3}\left[\text{D}\left(\frac{\phi_{\text{sw}}}{\phi}\right)^{2/\text{D}}+3-\text{D}\right]-3C-\frac{2L}{3} \ln \left[\left(\frac{N}{L}\right)^3 g^\text{D} \right]+2\alpha \ln \phi\Bigg\}, \label{eq:s-t-d} \\
&\frac{\sigma_{\text{RP}}}{(\Phi/4)n^\text{RP}_{x0}  k_B T} &&= \left[2C\frac{\phi_{\text{sw}}^{\frac{2}{\text{D}}-\frac{2}{3}}}{\phi^{\frac{2}{\text{D}}}}+L\left(\frac{\phi_{\text{sw}}^{\frac{1}{\text{D}}-\frac{1}{3}}}{\sqrt{g}\phi^{\frac{1}{\text{D}}}} -\frac{\sqrt{g}\phi^{\frac{1}{\text{D}}}}{\phi_{\text{sw}}^{\frac{1}{\text{D}}-\frac{1}{3}}} \right) -2\alpha\right]\phi, \label{eq:st-s-d} \\
&\frac{f_{\text{ABRP}}}{(\Phi/4)n^\text{ABRP}_{x0}  k_B T} &&= \phi\Bigg\{2CN\left[\beta\sqrt{\frac{\text{D}(\phi_{\text{sw}}/\phi)^{2/\text{D}} + 3-\text{D}}{3\phi_{\text{sw}}^{2/3}N}} -\ln \left(\frac{\sinh \beta}{\beta}\right)\right] +2\alpha \ln\phi +L\text{D}\left(\frac{\phi_{\text{sw}}^{\frac{1}{\text{D}}-\frac{1}{3}}}{\sqrt{g}\phi^{\frac{1}{\text{D}}}} +\frac{\sqrt{g}\phi^{\frac{1}{\text{D}}}}{\phi_{\text{sw}}^{\frac{1}{\text{D}}-\frac{1}{3}}} \right) \nonumber \\
&&&\indent +L(3-\text{D})\left(\frac{1}{\phi_{\text{sw}}^{\frac{1}{3}}}\sqrt{\frac{3-\text{D}}{3-\text{D}g}} + \phi_{\text{sw}}^{\frac{1}{3}}\sqrt{\frac{3-\text{D}g}{3-\text{D}}} \right) -\frac{2L}{3} \ln \left[\left(\frac{N}{L}\right)^3 g^\text{D} \right]\Bigg\}, \label{eq:abrp-d}\\
&\frac{\sigma_{\text{ABRP}}}{(\Phi/4)n^\text{ABRP}_{x0}  k_B T} &&= \left[\sqrt{\frac{N}{3\phi_{\text{sw}}^{2/3}}}\frac{2C\beta\left(\frac{\phi_{\text{sw}}}{\phi}\right)^{2/\text{D}}}{\sqrt{\text{D}(\phi_{\text{sw}}/\phi)^{2/\text{D}} + 3-\text{D}}}+L\left(\frac{\phi_{\text{sw}}^{\frac{1}{\text{D}}-\frac{1}{3}}}{\sqrt{g}\phi^{\frac{1}{\text{D}}}} -\frac{\sqrt{g}\phi^{\frac{1}{\text{D}}}}{\phi_{\text{sw}}^{\frac{1}{\text{D}}-\frac{1}{3}}} \right) -2\alpha\right]\phi, \label{eq:abrp-s-d} 
\end{align}
where $\beta$ satisfies Eq.~\eqref{eq:langevin-d}. Note that the crosslink density in each model can be related to the linear Young's modulus according to Sec.~III of the main text.